\documentclass{egpubl}
\usepackage{eg2023}
 
\ConferenceSubmission   
\usepackage[T1]{fontenc}
\usepackage{dfadobe}  

\usepackage{cite}  
\BibtexOrBiblatex
\electronicVersion
\PrintedOrElectronic
\ifpdf \usepackage[pdftex]{graphicx} \pdfcompresslevel=9
\else \usepackage[dvips]{graphicx} \fi

\usepackage{egweblnk} 
\usepackage{rotating}

\usepackage[normalem]{ulem}

\usepackage[flushleft]{threeparttable}
\usepackage{booktabs}
\usepackage{multirow}

\title[A Comprehensive Review of \\ \ Data-Driven Co-Speech Gesture Generation]%
      {A Comprehensive Review of \\ \ Data-Driven Co-Speech Gesture Generation}

\author[S. Nyatsanga \& T. Kucherenko \& C. Ahuja \& G. E. Henter \& M. Neff]
 {\parbox{\textwidth}{\centering S.~Nyatsanga$^{1}$ T.~Kucherenko$^{2}$ C.~Ahuja$^{3}$ G.~E.~Henter$^{4}$ M.~Neff$^{1}$
         }
         \\
{\parbox{\textwidth}{\centering $^1$University of California, Davis, USA\\        $^2$SEED - Electronic Arts, Stockholm, Sweden\\ $^3$Meta AI, USA \\ $^4$Division of Speech, Music and Hearing, KTH Royal Institute of Technology, Stockholm, Sweden \\
    }
  }
}
\usepackage{pifont} 
\usepackage[inline]{enumitem}
\newcommand{\cmark}{\ding{51}}

\newcommand{\new}[1]{{#1}} 

\begin{document}
\teaser{
 \vspace{-12mm}
 \includegraphics[scale=0.5]{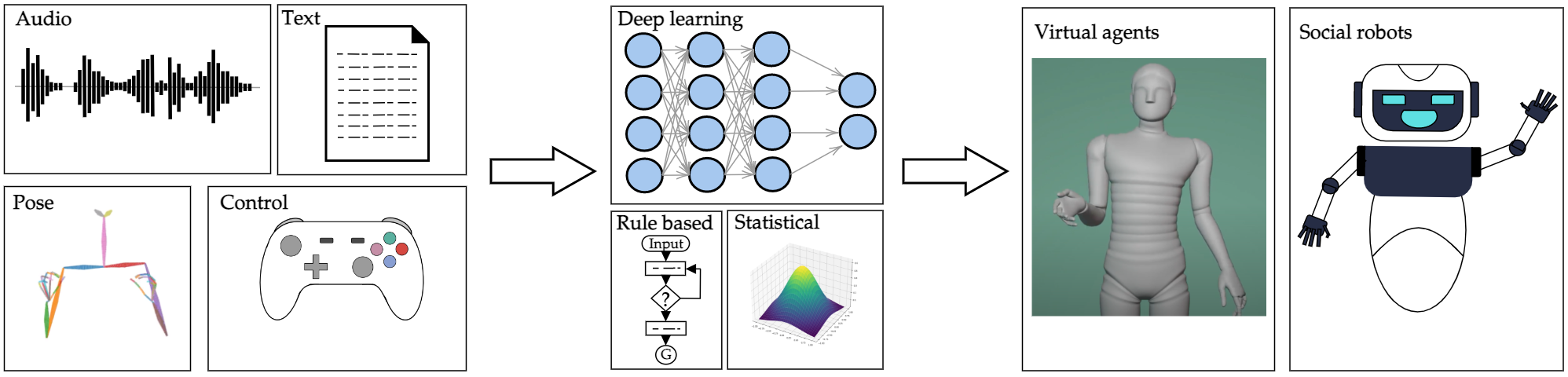}
 \centering
  \vspace{-3mm}
  \caption{Co-speech gesture generation approaches can be divided into rule-based and data-driven. Rule-based systems use carefully designed heuristics to associate speech with gesture (Section~\ref{sec:rule-based}). Data-driven approaches associate speech and gesture through statistical modeling (Section~\ref{subsec:statistical}), or by learning multimodal representations using deep generative models (Section~\ref{subsec:deep-learning}). The main input modalities are speech audio in an intermediate representation; text transcript of speech; humanoid pose in joint position or angle form; and control parameters for motion design intent. Virtual agents and social robotics are the main research applications, although also compatible with games and film VFX.}
  \vspace{1mm}
\label{fig:teaser}
}
\maketitle

\begin{abstract}

Gestures that accompany speech are an essential part of natural and efficient embodied human communication. The automatic generation of such co-speech gestures is a long-standing problem in computer animation and is considered an enabling technology for creating believable characters in film, games, and virtual social spaces, as well as for interaction with social robots. The problem is made challenging by the idiosyncratic and non-periodic nature of human co-speech gesture motion, and by the great diversity of communicative functions that gestures encompass.
The field of gesture generation has seen surging interest in the last few years, owing to the emergence of more and larger datasets of human gesture motion, combined with strides in deep-learning-based generative models that benefit from the growing availability of data. This review article summarizes co-speech gesture generation research, with a particular focus on deep generative models. First, we articulate the theory describing human gesticulation and how it complements speech. Next, we briefly discuss rule-based and classical statistical gesture synthesis, before delving into deep learning approaches. We employ the choice of input modalities as an organizing principle, examining systems that generate gestures from audio, text and non-linguistic input. Concurrent with the exposition of deep learning approaches, we chronicle the evolution of the related training data sets in terms of size, diversity, motion quality, and collection method (e.g., optical motion capture or pose estimation from video). Finally, we identify key research challenges in gesture generation, including data availability and quality; producing human-like motion; grounding the gesture in the co-occurring speech in interaction with other speakers, and in the environment; performing gesture evaluation; and integration of gesture synthesis into applications. We highlight recent approaches to tackling the various key challenges, as well as the limitations of these approaches, and point toward areas of future development.
   \\



\end{abstract}  
\section{Introduction}
This paper summarizes research on the synthesis of gesture motion, with a particular emphasis on more recent techniques using deep learning.  The focus is on {\em co-verbal gesture}, gesture that accompanies speech.  When considering the problem, a first reasonable question is ``Why should we care about gesture at all?''  Gesture plays at least three main functions.  First, and most simply, it helps artificial agents and robots look more alive and be more engaging (this has been shown multiple times, e.g.~\cite{salem2012generation,salem2013err, breazeal2016social,saunderson2019robots}).  Second, it communicates functional information.  This can include pointing or deictic gesture that establish reference; emblems that replace words, and imagistic metaphoric and iconic gestures that illustrate concepts and artifacts. Third, gesture communicates social information, including personality~\cite{smith2017understanding,neff2010evaluating,neff2011don,durupinar2016perform}, emotion~\cite{volkova2014emotion,xu2013mood,giraud2015perception,normoyle2013effect,dael2013perceived,fourati2016perception,castillo2019we} and subtext.  

\new{Before summarizing work on gesture synthesis, it is worthwhile to consider how gesture can support a range of applications for virtual agents and robots.}
First, it is well established that gestures do indeed communicate~\cite{goldin2005hearing,kendon1994gestures,hostetter2011gestures,goldin1999role} 
Hostetter’s meta-analysis~\cite{hostetter2011gestures} presents three main findings for when gestures communicate: gestures depicting motor actions are more communicative than those depicting abstract topics; gestures that are not completely redundant have a larger impact on communication, and children benefit more from gesture than adults.

\new{Gestures communicate in a different manner} than spoken language. They communicate particularly directly when being used to describe spatial concepts or object manipulation because there is a natural iconicity to these concepts, which is well portrayed in  gestures. ``Gesture permits the speaker to represent ideas that are compatible with its mimetic and analog format (e.g. shapes, sizes, spatial relationships) - ideas that may be less compatible with the discrete and categorical format underlying speech. Thus, when it accompanies speech, gesture allows speakers to convey thoughts that may not easily fit into the categorical system that their spoken language offers.''~\cite{goldinandmcneill1999role}. The iconicity of gestures makes them more transparent than language, which is purely symbolic.  Tversky argues that ``[gestures] take advantage of human ability to make rapid spatial judgments and inferences. Neither depictions nor gestures can convey all the information surrounding an idea or set of ideas; this forces them to extract what is essential, to simplify the ideas, making them easier to comprehend and remember.'' ~\cite{tversky2007communicating}.  They provide an additional code, a motor code for information, and additional codes are known to improve memory.  Gestures are particularly congruent with actions or transformations~\cite{tversky2007communicating}. \new{A more detailed gesture typology is presented in Section~\ref{sec:humanGest}.}

Nonverbal communication appears to be particularly important in providing appropriate social cueing~\cite{whittaker2003theories}. Feelings, emotions and attitudes are often not made verbally explicit and must be inferred from nonverbal channels. The presence of nonverbal communication can radically change the outcome of an exchange.  For example, a study comparing face-to-face and voice only union negotiations showed greater interpersonal communication in the face-to-face setting, whereas the speech-only communication focused more on content, was more impersonal, saw reduced praise, greater blame, more disagreement and was more likely to end in deadlock~\cite{rutter1981visual}. 

Additionally, gestures can be used to regulate a dyadic or group interaction by managing turn-taking \cite{whittaker2003theories,kipp2005gesture}. Kipp defines turn-taking as ``assigning, holding or yielding a turn'' in a dialog \cite{kipp2005gesture}. Bavelas \cite{bavelas1994gestures} identified so-called ``turn gestures'' in dialogue interactions, with sub-categories of gestures that indicate: giving turn to the other speaker, accepting a turn from the other speaker or offering turn to any speaker in a group.

\new{Gesture can also support particular applications, for example, t}here is a growing body of evidence showing that gestures help people learn. There are at least three mechanisms by which this happens: learners watching a teacher gesture, learners performing gestures themselves, and teachers adapting their instruction based on information gained from the learner's gestures.  For an excellent summary, see~\cite{novack2015learning}.  Gestures can link abstract concepts in the immediate environment, reduce cognitive load and enhance spoken communication~\cite{novack2015learning}.  A recent meta-analysis~\cite{davis2018impact} of twenty experiments showed that the gesture already present in pedagogical agents is beneficial for student learning, with positive effects on transfer of knowledge, retention of learning and agent persona, but not on reducing cognitive load.

Given the potential value of nonverbal communication, the question remains as to how to synthesize appropriate behavior in our computational systems.  Formally, the problem is to generate an input to output mapping, where the input is some representation of the content \new{to be expressed} and the output is some representation of the behavior \new{to perform}.  Most learning systems to date have assumed audio and/or text as the input, although we will see this has limitations.  The output is generally either frames of pose data or a lexicalized representation of the gestures that should appear (e.g. accompany this sentence with a conduit gesture that displays the idea of conveying something to the interlocutor).  The latter must then be converted to animation using some secondary system.

Previous surveys have covered co-speech gesture synthesis with varying scope and emphasis \cite{wei2022learning,liu2021speech,ye2021human}. Wei et al.~\cite{wei2022learning} focus on audio-visual learning for human-like machine perception, and briefly cover gesture synthesis.
We focus on co-verbal gesture synthesis, including its theory, synthesis techniques with an emphasis on deep learning, and an eye towards application to virtual agents and robots. Ye et al.~\cite{ye2021human} surveyed deep-learning-based human motion modeling, and thus is related to our work via the emphasis on deep generative models.
Our survey covers a larger scope of deep-learning-based gesture synthesis research and offers a more comprehensive set of key challenges that are specific to the problem. The survey by Liu et al.~\cite{liu2021speech} is more closely related to our work, emphasizing co-verbal gesture generation for virtual agents and robots. Their work presents a scoping review of data-driven techniques for speech-to-gesture generation, related datasets, and evaluation metrics. We include a larger set of papers (40 vs. 19) and are able to provide a more in depth treatment of the technical material given the substantially longer STAR format.
\begin{figure*}[h]
    \centering
    \includegraphics[scale=0.5]{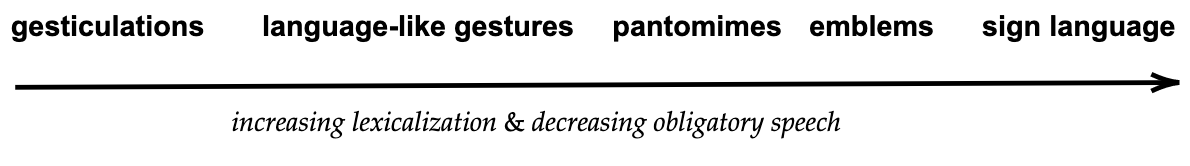}
    \caption{Kendon's Continuum of gesture categories, as described by McNeil \cite{mcneil1992handmind,mcneill2008gesture}}
    \label{fig:kendonscont}
\end{figure*}

Overall, our survey makes the following contributions to the field:
\begin{itemize}
    \item A detailed discussion on the theory and motivation for co-verbal gesture synthesis.
    \item A discussion on rule-based and statistical techniques, illustrating how these approaches can complement the strengths and weaknesses of recent deep-learning approaches.
    \item An emphasis on deep-learning-based generation systems using input modality as an organizing principle for the research. 
    \item A discussion on the most commonly used speech-to-gesture datasets, collected via motion capture or pose estimation.
    \item Identifying and detailing a set of key challenges for co-verbal gesture synthesis and potential research directions.
\end{itemize}

The remainder of the paper begins by providing a deeper background on gesture, followed by a summary of synthesis techniques that have been developed to date and concludes with a discussion of major open problems.


\section{Human gesticulation}
\label{sec:humanGest}


Manual Gestures are non-verbal, non-manipulative hand/arm movements that occur during speech \cite{mcneil1992handmind, tuite1993production, kendon1983gestureandspeech}. We will refer to manual gestures as simply ``gestures'' in this work, although gestures can in general be performed by other body parts, such as the head. Gestures aid in the communicative intent and are closely linked to accompanying speech in terms of timing, meaning and communicative function. For instance, gestures can be used for pointing to resolve references to objects (``what is that'') or illustrate concepts that would otherwise be difficult to explain verbally \cite{kipp2005gesture}. Therefore, gestures play an important complementary role to speech because they enable broader and more efficient expression of personality, emotion and motivation of the speaker \cite{ekman1969nonverbal,mehrabian1968some,ekman1992argument}. Additionally, gesture plays an important cultural role, because members of a community can either identify with or easily understand the emotions and attitudes of those around them through these non-verbal cues \cite{borkenau1993consensus}.

Gestures can take many forms depending on the speaker, and the morphological rules governing their construction equally vary. Based on Kendon's gesture categorization \cite{kendon1988gestures}, McNeill proposed ``Kendon's Continuum'' \cite{mcneil1992handmind, mcneill2008gesture} where gesture categories are sorted in increasing \textit{lexicalization}, that is the degree to which they adhere to formal, language-like grammatical rules, as illustrated in Figure \ref{fig:kendonscont}. In this framework, the least lexicalized (conversational gesture) have obligatory speech, while the fully lexicalized (sign language) have little or no obligatory speech as the gestures themselves gain explicit lexical properties.  Most importantly, the fully lexicalized end of the spectrum (sign languages) have formal syntactic structure like spoken languages, but this is absent from coverbal gesticulations.  This lack of structure is one of the challenges in producing coverbal gesture as the behavior can be highly idiosyncratic.

\begin{figure*}
    \centering
    \includegraphics[scale=0.4]{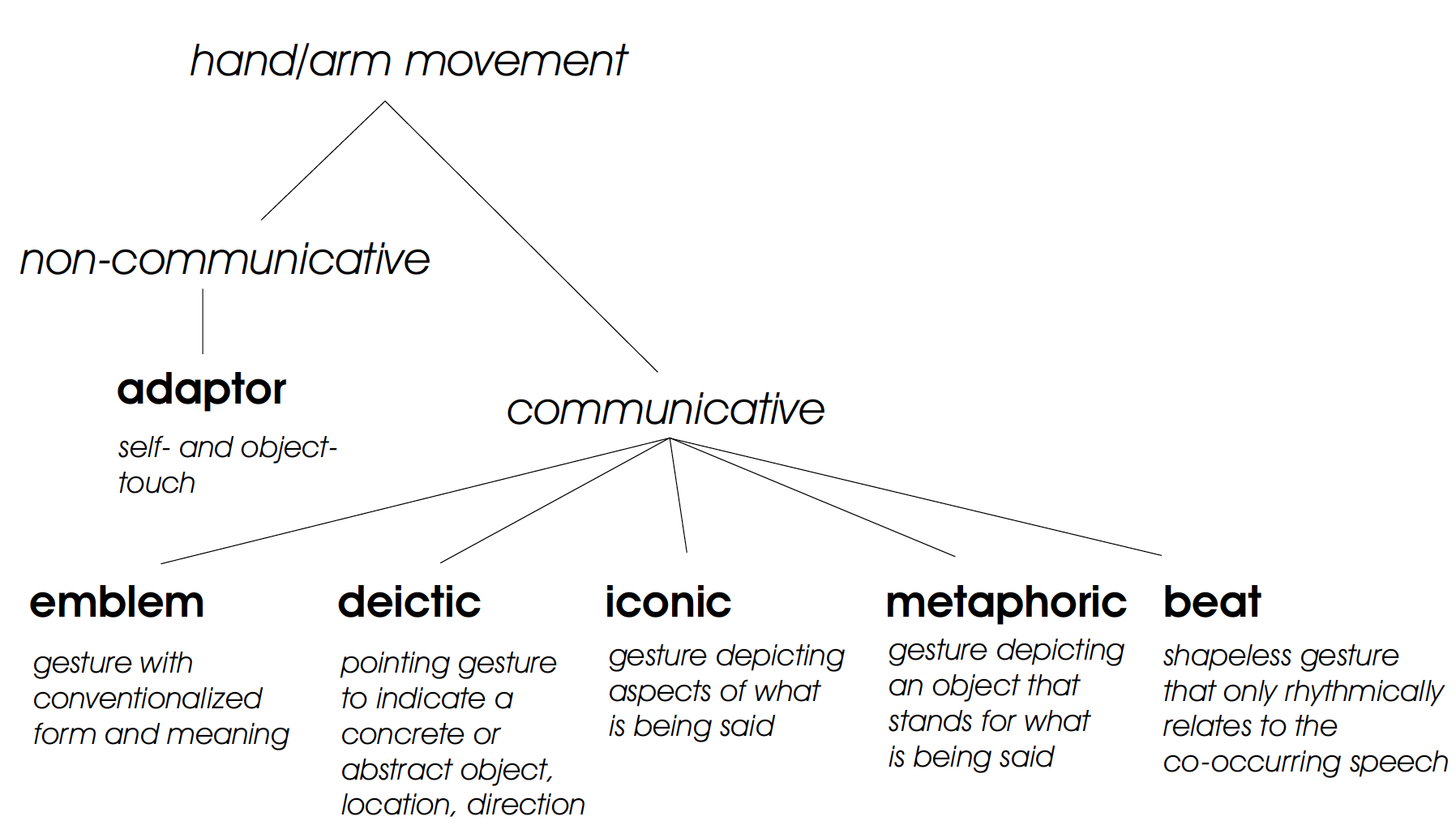}
    \caption{Relational graph of gesture categories and their defining properties. Figure from Kipp \cite{kipp2005gesture} (used with permission)}
    \label{fig:gesturetree}
\end{figure*}

There are many different forms of gesture and McNeill~\cite{mcneill1992hand,mcneill2008gesture} argues for a dimensional view in which the dimensions are iconic (images of the concrete), metaphoric (images of the abstract), deictic (pointing) and beat (flicks of the hand in time to rhythm of the speech). See Figure~\ref{fig:gesturetree}.  An iconic gesture might show the size of a box being discussed by drawing it in space, whereas a metaphoric gesture might indicate an abstract concept, such as all ideas are included, by making an umbrella shape.  A given gesture may load on multiple of these dimensions, for example displaying both iconicity and deixis. An additional category are adaptors (or self-adaptors), which are self manipulations such as scratching one's nose or bracing fingers.  These are not designed to communicate, but do convey information about personality~\cite{neff2011don}.  

Kendon introduced a three-level hierarchy to describe the structure of gestures~\cite{kendon1972some}. The top level is the {\em gesture unit}.  Gesture units start in a rest pose, contain a series of gestures, and then return to a rest pose.  The starting and ending rest pose need not be the same.  A {\em gesture phrase} encapsulates an individual gesture in this sequence.  Each phrase can in turn be broken down into a sequence of {\em gesture phases}.  These include:  a {\em stroke}, which is the main meaning carrying movement of the gesture and has the most focused energy; a {\em preparation}, which is a motion that takes the hands to the required position and orientation for the start of the stroke;  a {\em prestroke hold}, which is a period of stillness in which the hands are held at the staring point of the stroke, before the stroke begins; a {\em poststroke hold}, in which the hands are held at the end position of the stroke; and finally, a {\em retraction}, that returns the hands to a rest pose.  All phases are optional except the stroke.  The pre- and poststroke holds function to synchronize the gesture with speech.

There are many challenges for automatically synthesizing gesture, for instance to drive virtual agents in human-computer interaction. One theory on the origins of gesture, the growth point hypothesis~\cite{mcneil1992handmind}, argues that gesture and language emerge from a common communicative intent.  Some communication may take verbal form and some nonverbal, with some being replicated across both.  Some agent architectures, such as SAIBA~\cite{kopp2006towards}, have tried to model this communicative intent.  This allows nonverbal communication to be unique, carrying different information than the verbal channel.  Many gesture synthesis approaches, and all deep learning approaches that we are aware of, do not model a communicative intent.  Instead, they synthesize gesture from audio, text or both.  This necessarily means that the gestures will be redundant with these other channels, and thus more limited than actual human gesture.

Another challenge is that gesture is idiosyncratic~\cite{mcneill2000language}, so different people may gesture in very different manners.  The same person may also generate different gesture even when delivering the same text. Finally, gestures are synchronized in time with their co-expressive speech. About 90\% of the time, the gesture occurs slightly before the co-expressive speech~\cite{nobe2000most} and rarely occurs after~\cite{kendon1972some}. While the earlier occurrence of gesture is common in human behavior, and research on animated characters also indicates a preference for this slightly earlier timing,it also indicated that people may not be particularly sensitive to errors in timing within a +/- .6 seconds~\cite{wang2013influence}.

\section{Approaches for gesture synthesis}

Synthesizing co-speech gestures is essential for creating interactive and believable virtual characters in human-computer interaction (HCI), graphics and social robotics. Thus significant effort has been applied and progress has been made for applications in virtual agents \cite{cassell2001beat,pelachaud2003computational,kopp2004synthesizing,neff2008gesture,chiu2015predicting} and humanoid robots \cite{salem2009towards,yoon2019robots,le2011design,go2018andorid,ng2010synchronized,holladay2016rogue}. Neff identifies the two main sub-problems of generating gesture as the \textit{specification problem} and the \textit{animation problem} \cite{neff2016hand}. The specification problem is concerned with determining \textit{what} gestures are to be performed by the character, and the animation problem entails \textit{how} to generate the appropriate hand motion. Gesture specification can use a range of inputs including speech, prosody, text and communicative intent, where rule-based, statistical and learning-based models have been used to determine the appropriate gesture. Similarly, gesture animation has used a range of procedural, physics-based or learning-based models to produce the hand motion.

Gesture generation models can be divided into two main categories: rule-based and data-driven. Within the latter, there are two sub-categories, statistical and learning-based. Rule-based systems \cite{cassell2001beat,pelachaud2003computational,kopp2004synthesizing,salem2009towards,cassell1994animated,kopp2002model,pelachaud2002embodied,lee2006nonverbal,thiebaux2008smartbody,marsella2013virtual} use carefully designed heuristics to select the appropriate gesture for a given speech input. Data-driven systems instead learn to associate speech with corresponding gestures from the data, and we expound on the sub-categories next. Statistical systems \cite{neff2008gesture,kipp2005gesture,bergmann2009gnetic,yang2020statistics} typically precompute probabilities or assign a prior distribution over the given gesticulation data and a gesture is sampled from the distribution based on speech input. Learning-based models \cite{chiu2015predicting,yoon2019robots,levine2010gesture,chiu2011train,hasegawa2018evaluation,ahuja2019react,ferstl2018investigating,kucherenko2019analyzing,ginosar2019learning,ahuja2020style,alexanderson2020style,kucherenko2020gesticulator,li2021audio2gestures,ferstl2019multi} make the fewest assumptions about the distribution of gesticulations and instead optimize the parameters of a complex non-linear function to map the input speech into the appropriate gesture. This non-linear function is usually implemented as a deep neural network with some form of a gradient-based optimization algorithm, and thus we simply refer to them as deep learning approaches. While animation may be synthesized using a range of methods for each technique, rule-based and statistical approaches have generally predicted a gesture label that is used to index either hand-animated or pre-recorded gesture clips that are then used to synthesize the final sequence. In contrast, deep-learning approaches have tended to synthesize motion on a per-frame basis.

\begin{figure}
    \centering
    \includegraphics[scale=0.3]{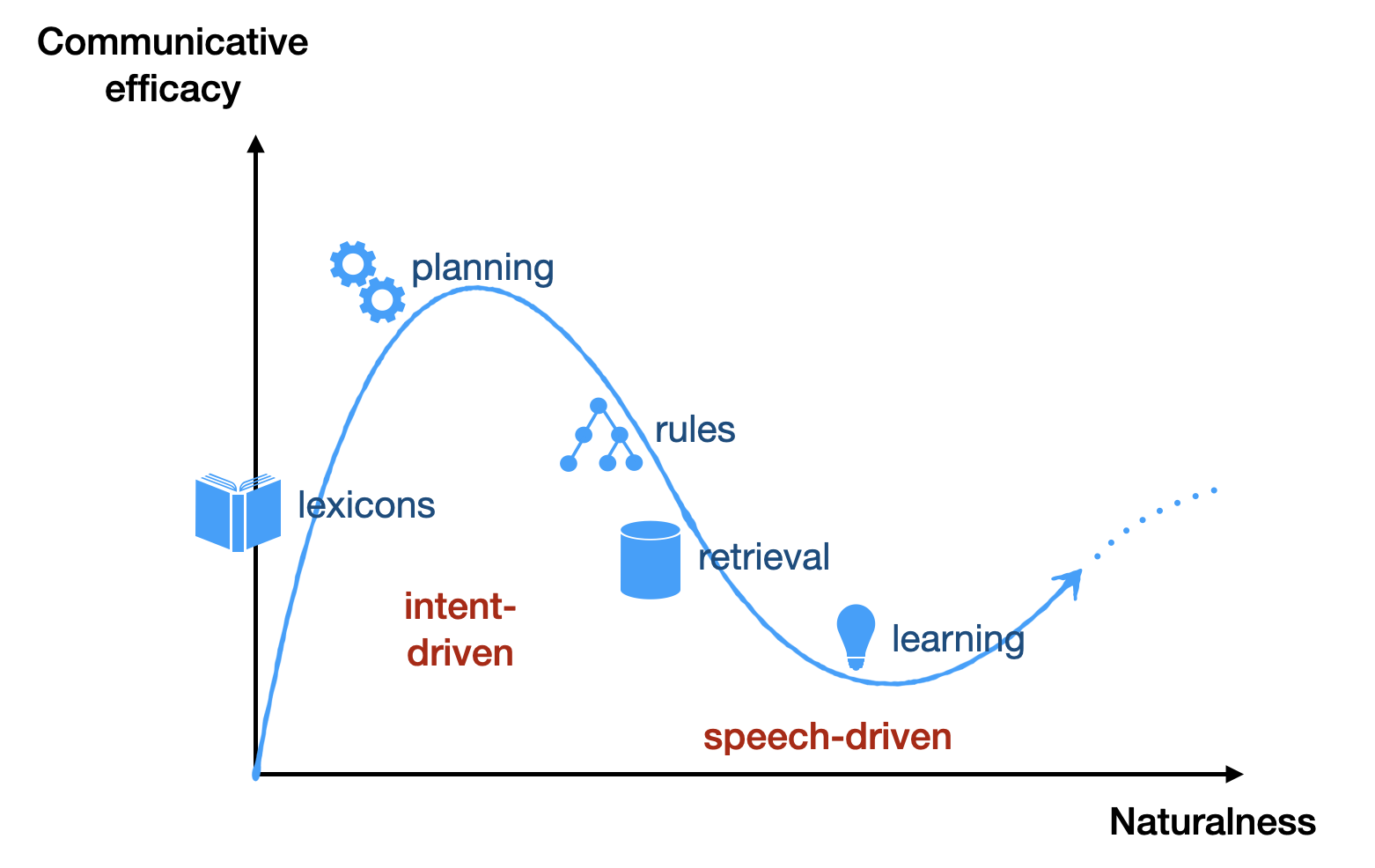}
    \caption{Overview of the development of the gesture generation field, as outlined by Stefan Kopp at the GENEA Workshop 2020. Figure by Stefan Kopp. Used with permission.}
    \label{fig:stefan_overview}
\end{figure}

Figure \ref{fig:stefan_overview} illustrates the development of the gesture generation field, specifically how different approaches handle the trade-off between naturalness and communicative efficacy. The early approaches were intent-driven and hence had high communicative efficacy \cite{cassell2001beat, kopp2004synthesizing, thiebaux2008smartbody}. They were not very natural, since they were mainly inserting pre-defined animations. Later approaches used statistics to analyze and retrieve gestures from large databases\cite{neff2008gesture,kipp2005gesture,bergmann2009gnetic}. Statistical approaches improved gesture naturalness, while slightly compromising communicative efficacy. Finally, modern approaches are mainly deep-learning-based, making the fewest assumptions about the underlying distribution of gesture data \cite{kucherenko2021moving, ahuja2020style, yoon2020speech}. Deep learning-based approaches can generate continuous and fairly natural gestures, but they are significantly less communicative. Motivated by this challenging trade-off, recent notable research has proposed hybrid systems for generating natural and semantically meaningful gestures, by combining rule-based and deep learning-based approaches. \cite{kucherenko2022multimodal, zhou2022gesturemaster, habibie2022motion}. We review the seminal approaches in rule-based, statistical and learning-based generation next. \new{In Section \ref{sec:rule-based}, we discuss what, in our estimation, are some of the most impactful, rule-based systems. We discuss them in chronological order for ease of understanding and to emphasize how the approaches influenced one another. The selected works have in common that they pioneered approaches for speech-driven hand or facial animation by devising heuristics and domain-specific languages for modeling behavior intent, planning, and realization. Since the focus of the paper is on data-driven systems, we only review these selected works. For a more detailed review of rule-based systems, we recommend the review article by Wagner et al.\cite{wagner2014gesture}.}

\section{Rule-based approaches} \label{sec:rule-based}

Cassell et al.~\cite{cassell1994animated} presented \textit{Animated Conversation}, the first rule-based system to automatically generate context-appropriate hand gestures, facial movements and intonation patterns between multiple human-like agents. Notably, this work was one of the first to explore the latent relationship between speech and gesture for generating realistic animation. The system initiated a dyadic interaction between two agents through a dialogue generator and planner. The generated text was transformed into speech through a text-to-speech system \cite{liberman1985structure} and deep symbolic representations were used to encode timing, intonation and the corresponding gesture prototypes. The gesture prototypes were used to perform the full gesture. The result was agents with appropriate and well-synchronized speech, intonation, facial expressions and hand gestures. However, the system was limited to domain-specific dialogue generation between two agents, which not only restricted free-form conversation (by restricting the discourse) and gesture animation, but also precluded real-time interaction with a human user.

Th\'{o}rrison proposed \textit{Ymir}, \cite{thorisson1996communicative} which improved on the Animated Conversation framework by enabling multimodal input from a user, including speech, gaze, gesture and intonation. It consisted of multiple modules for input perception, dialogue generation, decision making and action schedulers in order to produce well-synchronized hand animation. However, although this offered more interactivity with a user, the system could only produce limited multi-modal output in real time. The work of Cassell et al.~\cite{cassell2000human} subsequently improved on the two frameworks by integrating the real-time multi-modal interactivity of Ymir with the symbolic generation and richer multi-modal synthesis capability of Animated Conversation. The result was an embodied conversational agent framework that produced reactive characters that behaved intuitively and robustly in conversations, albeit still limited to dialogue deriving from a static knowledge base.

Another of the seminal works in a rule-based generation was the Behaviour Expression Animation Toolkit proposed by Cassell et al.~\cite{cassell2001beat}. BEAT took typed text as input and could synthesize well-synchronized speech, gesture, facial animation and intonation. The system used contextual information latent in the text to choose pre-recorded hand, arm and facial movements by relying on a set of carefully designed heuristics from previous nonverbal conversational behavior research. BEAT was highly extensible allowing animators to insert new rules that parameterize personality, movement characteristics, scene constraints and desired animation style.

Alternatively, Kopp et al.~\cite{kopp2002model}  proposed a model-based approach for generating complex multimodal utterances (i.e. speech and gesture) from XML specifications of their form. Instead of relying on pre-recorded gestures, \new{as the previously discussed approaches did}, the system applied non-uniform cubic B-Splines to form a gesture trajectory that satisfies all velocity and position constraints. The authors demonstrated the multimodal capabilities of the system through Max: a virtual reality-based agent that interacts and assists a human user through construction tasks, by using prosodic speech, deictic and iconic gestures, gaze and emotive facial expressions \cite{kopp2003max}.

Facial expressions, gaze direction and head movements are essential non-verbal behaviors that communicate the intent and emotional state of a speaker. They can also act as facial gestures e.g. ``raised eyebrow'' or gaze direction utilized to resolve a referent object or direction. Therefore, endowing virtual agents with such qualities can make them more anthropomorphic. Pelechaud et al.~\cite{pelachaud2002embodied}, developed Greta: a 3D virtual agent whose facial gestures communicated the agent's emotional state. The system was designed as a BDI agent (i.e. prior Beliefs, Desires and Intentions) \cite{rao1991modeling}. A Dynamic Belief Network (DBN) modelled Greta's constantly evolving emotions and computed the triggering thresholds and evolution of her emotions, resulting in emotive verbal and non-verbal behaviour.

The development of new rule-based systems often necessitated the development of a new domain-specific language (DSL), usually based on XML. Examples of these include an XML processing pipeline in the BEAT system \cite{cassell2001beat}, MURML for multimodal behavioral planning and animation of Max \cite{kopp2003max, kranstedt2002murml}, APML for representing the agent's behaviour semantically \cite{carolis2004apml}, and RRL for representing simulations of multimodal dialogue \cite{piwek2004rrl}. However, these DSLs were often incompatible with each other even as the systems solved similar or overlapping objectives. As a result, a group of researchers developed a unified language for multimodal behaviour generation for virtual agents, called the Behavior Markup Language (BML) \cite{kopp2006towards,vilhjalmsson2007behavior}. BML was designed in the context of a comprehensive framework with intent planning, behavior planning and behavior realization stages. Within this framework, BML described the desired physical realization and thus connected behavior planning to behavior realization. BML became the standard format for rule-based systems, finding use in open-source frameworks like SmartBody \cite{thiebaux2008smartbody}, and other agent embodiments like humanoid robots \cite{le2011design}.

The development of BML led to continued advances in rule-based systems, even as some research started to explore learning-based systems. For instance, Marsella et al.~\cite{marsella2013virtual} generated facial expressions and behaviors (including gestures, head movements, eye saccades, blinks and gaze), for a 3D virtual character, by analyzing the prosodic and acoustic features of speech, as well as shallow analysis of the utterance text to determine rhetorical and pragmatic content. Ravenet et al.~\cite{ravenet2018automating} generated metaphorical gestures by leveraging BML to extract metaphorical properties from input speech. \new{Their system leveraged BML annotations to synchronize speech audio and gestures, and configure gesture characteristics (e.g., hand shape, movement, orientation) to convey the desired representational meaning during behavior realization. Overall, BML continues to be a standard domain-specific language for behavior planning and realization in rule-based gesture generation systems.}


Rule-based gesture generation systems can produce high-quality gestures that are well synchronized with speech. Due to their reliance on pre-recorded motion, hand animation or carefully engineered systems for generating gestures, rule-based systems can have better motion quality than learning-based systems.  Hand-tuned rules may also better preserve semantics within their limited domain. However, the gesture distribution is often not diverse. Moreover, the carefully designed rules require significant expert knowledge which is laborious and not scalable. Such systems are inflexible in that they can only produce a small set of plausible gestures for a particular speech input or scenario. Therefore, the inability to produce diverse gestures in a non-deterministic manner means the resulting virtual agents (or any other embodiments) can only behave in an expressive and naturalistic way for limited examples. Data-driven methods were proposed to try to overcome these limitations. Given the overall advances in deep learning, they may eventually also produce the highest quality motion. We review the two data-driven sub-categories next, statistical and deep learning-based methods.

\section{Data-driven approaches} \label{sec:data-driven}

\subsection{Speech-Gesture Datasets}
Any data-driven method is fundamentally limited by the data it is trained on.
The number of datasets suitable for machine-learning on human gesture data has been steadily rising, as has their size.
Table~\ref{table:datasets} provides an overview of major datasets for gesture generation, and their characteristics such as size, motion format (2D or 3D), modalities, included annotation, and more.
It is seen that the dataset sizes have reached new heights of 100+ hours in recent years, and there is also greater diversity in terms of the number of speakers, thanks to 3D pose estimation from video. Unfortunately, only a small fraction of datasets contain high-quality finger motion, which is of great importance for generating expressive and meaningful gestures.



There are two main methods for obtaining motion data for gesture synthesis: optical motion capture \cite{metallinou2016usc, takeuchi2017creating, lee2019talking, joo2019towards, ferstl2018investigating} or pose estimation from monocular video \cite{yoon2019robots, ahuja2020no, jonell2019learning, kucherenko2022multimodal,habibie2021learning}. 

Existing datasets recorded using motion capture are usually smaller, since that method of data collection is much more expensive and labor-intensive, and generally takes place in a controlled studio environment. Emotion is often acted. The main advantages of the resulting data is that movements are in 3D and have high quality. This method is also the best at capturing finger motion.  


Datasets instead obtained from pose estimation can be an order of magnitude larger, as they can be sourced from online videos. This enables finding genuine, in-the-wild gestures, and the material can be large enough to include much more diversity. The downsides are relatively lower motion quality (fingers being especially hard) and being limited to 2D motion only. Recent monocular video work has lifted the skeleton motion to 3D~\cite{habibie2021learning}.


In practice, the amount of data needed is likely to depend on the application at hand. While gathering data from a specific target speaker of interest is usually better than having an equivalent amount of data from non-target speakers, the gesture manifolds of different speakers nonetheless often have a significant overlap. It has been found that, starting from a generative model trained on one individual style, one requires only two minutes of data to fine-tune a gesture generation model for another style \cite{ahuja2022low}.
Recent work has also demonstrated the possibility of learning to embed different gesture styles, which then can be used for zero-shot adaptation to the style of an unseen target speaker with no training data of the target speaker \cite{fares2022zero, ghorbani2022zeroeggs}.
Techniques for augmenting gesture data so as to increase the amount of motion data for training have also been studied, especially mirroring \cite{windle2022pose, windle2022arm}.


\begin{table*}[]
\tiny
\setlength\tabcolsep{1.5pt} 
\begin{tabular}{|l|l|l|l|l|l|l|l|}
\hline
\textbf{Name}                             & \textbf{Size}                                               & \textbf{\# of sp.} & \textbf{Mot. format} & \textbf{Modalities}                                           & \textbf{HQ fing.} & \textbf{Dialog?}     & \textbf{Link}                                                                                       \\ \hline
IEMOCAP    \cite{busso2008iemocap}                               & 12h   & 10 & mp4 video         & Ges., Audio, Text, Emotion       &   &  Dialog &  \href{https://sail.usc.edu/iemocap/}{sail.usc.edu/iemocap/}     \\ \hline

SaGA     \cite{lucking2013data}        & 1 h     & 6     & mp4 video              & Ges., Audio, Gest. properties                           &                             & Dialog                        &  \href {https://www.phonetik.uni-muenchen.de/Bas/BasSaGAeng.html}  {www.phonetik.uni-muenchen.de/Bas/BasSaGAeng.html}               \\ \hline
Creative-IT \cite{metallinou2016usc}                              &      2h                                                       &         16                    &   3d joint rot.                     & Ges., Audio, Text, Emotion                                       &   & Dialog & \href{https://sail.usc.edu/CreativeIT/ImprovRelease.htm}{sail.usc.edu/CreativeIT/ImprovRelease.htm}                      \\ \hline

MPI-EBEDB \cite{volkova2014mpi} & 1.43h  & 8 & 3d joint rot. & Ges., Text & & Monolog & \href{http://ebmdb.tuebingen.mpg.de/}{ebmdb.tuebingen.mpg.de} \\
\hline

Gesture-Speech Dataset  \cite{takeuchi2017creating}    &      5h                                                       &         2                    &   3d joint rot.                     & Ges., Audio                                      &  \cmark & Monolog & \href{https://bit.ly/2Q4vSnT}{bit.ly/2Q4vSnT}                      \\ \hline

CMU Panoptic   \cite{Joo_2017_TPAMI}                           &       5.5 h                                                      &  50 &       3d joint rot.                 &         Ges., Audio, Text                                                      &   & Dialog & \href{http://domedb.perception.cs.cmu.edu/}{domedb.perception.cs.cmu.edu/}                                   \\ \hline

Trinity Speech-Gesture I   \cite{ferstl2018investigating}         & 6 h & 1     & 3d joint rot.     & Ges., Audio                                              &                             & Monolog                       & \href{https://trinityspeechgesture.scss.tcd.ie/data/Trinity\%20Speech-Gesture\%20I/}{trinityspeechgesture.scss.tcd.ie/data/TrinitySpeech\-GestureI/}    \\   \hline

Speech-Gesture \cite{ginosar2019learning}                           & 144 h                                                   & 10     & 2d coords.         & Ges., Audio                                               &                             & Monolog                       & \href{https://github.com/amirbar/speech2gesture/blob/master/data/dataset.md}{github.com/amirbar/speech2gesture}  \\ \hline
TED Dataset     \cite{yoon2019robots}                          & 106 h                                                       & 1,295  & 2d coords.         & Ges., Audio                                              &                             & Monolog                       & \href{https://github.com/youngwoo-yoon/youtube-gesture-dataset}{github.com/youngwoo-yoon/youtube-gesture-dataset}               \\ \hline
Talking With Hands  \cite{lee2019talking}    & 50h    &   50    & 3d joint rot.     & Ges., Audio     & \cmark                           & Dialog   & \href{https://github.com/facebookresearch/TalkingWithHands32M}{github.com/facebookresearch/TalkingWithHands32M}                \\ \hline
PATS  \cite{ahuja2020no}                                    & 250 h                                                  & 25     & 2d coords.         & Ges., Audio, Text                                         &                             & Monolog                       & \href{http://chahuja.com/pats}{chahuja.com/pats}                                                \\ \hline
Trinity Speech-Gesture I  &  & & & & & & \\ GENEA Extension   \cite{kucherenko2021large}        & 6 h & 1      & 3d joint rot.     & Ges., Audio, Text                                              &                             & Monolog                       &         \href{https://trinityspeechgesture.scss.tcd.ie/data/Trinity\%20Speech-Gesture\%20I/GENEA_Challenge_2020_data_release/}{TrinitySpeech\-GestureI/GENEA\_Challenge\_2020\_data\_release/}                      \\ \hline
Trinity Speech-Gesture II   \cite{ferstl2021expressgesture}         & 4 h & 1      & 3d joint rot.     & Ges., Audio, Gest. segment.                                              &                             & Monolog                       & \href{https://trinityspeechgesture.scss.tcd.ie/data/Trinity\%20Speech-Gesture\%20II/}{trinityspeechgesture.scss.tcd.ie/data/TrinitySpeech\-GestureII}                               \\ \hline

Speech-Gesture 3D extension   \cite{habibie2021learning}                        & 144 h                                                   & 10     & 3d coords.         & Ges., Audio                                               &                             & Monolog                       & \href{https://nextcloud.mpi-klsb.mpg.de/index.php/s/7LzxGSepzrndg2x}{nextcloud.mpi-klsb.mpg.de/index.php/s/7LzxGSepzrndg2x}    \\ \hline

Talking With Hands &  & & & & & & \\GENEA Extension \cite{yoon2022genea} & 20h    &         17     & 3d joint rot.     & Ges., Audio, Text                                     &  \cmark                          & Dialog                        & \href{https://zenodo.org/record/6998231}{zenodo.org/record/6998231}                                      \\ \hline


SaGA++              \cite{kucherenko2022multimodal}                      & 4 h                                                     & 25    & 3d joint rot.     & Ges., Audio, Gest. properties                           &                             & Dialog                        & \href{https://svito-zar.github.io/speech2properties2gestures/}{svito-zar.github.io/speech2properties2gestures}                                      \\ \hline

ZEGGS Dataset \cite{ghorbani2022zeroeggs} & 2 h & 1 & 3d joint rot. & Ges., Audio & \cmark & Monolog & \href{https://github.com/ubisoft/ubisoft-laforge-ZeroEGGS}{github.com/ubisoft/ubisoft-laforge-ZeroEGGS} \\ \hline

BEAT Dataset \cite{liu2022beat} & 76 h & 30 & 3d joint rot. & Ges., Audio, Text, Gest. properties, Emotion & \cmark & Both & \href{https://pantomatrix.github.io/BEAT/}{pantomatrix.github.io/BEAT} \\ \hline
\end{tabular}
\caption{Speech-gesture datasets ordered from oldest to newest. The following abbreviations were used: ``sp.'' for ``speakers'', ``mot. format'' for ``motion format'', ``HQ fing.'' for ``high-quality fingers``, ``coords.'' for ``coordinates'', and ``rot.'' for ``rotations''.}
\label{table:datasets}
\end{table*}



\subsection{Statistical and early machine learning approaches} \label{subsec:statistical}

In statistical systems, the latent relationship between speech and gesture is modeled by the statistics of the underlying gesture distribution, instead of being encoded by an expert. Compared to rule-based systems, statistical approaches make fewer assumptions about the speech-gesture association, instead either pre-computing \new{conditional} probabilities for the gesture data or assigning a prior probability distribution. \new{Similar to our approach in Section \ref{sec:rule-based}, we focus on a subset of statistical approaches that, in our estimation, are some of the most impactful in the field. The works are described in chronological order to illustrate advances in statistical systems.}

Kipp \new{proposed one of the earliest statistical systems, which} modeled an individual's gesture by analyzing an annotated co-speech dataset and producing a \textit{gesture profile} \cite{kipp2005gesture}. The data was annotated using the video annotation tool ANVIL \cite{kipp2001anvil} to define a gesture profile consisting of individual properties such as handedness, timing and communicative function. The gesture profiles were then modeled using statistical models inspired by work in speech recognition and dialogue act recognition \cite{reithinger1997dialogue}. The plausibility of a gesture was estimated using conditional probabilities on gesture bi-grams, and the occurrence of the gesture given semantics from input text. The result was statistical models for an individual's gesture properties like handedness, transitions and timing, forming an individual's gesture profile. The profiles were then used to generate plausible gestures from annotated input speech. The generation process had distinct stages: 1) Assigning semantic tags to input text; 2) Generating all possible gestures, adding them to an intermediate graph representation, and assigning probability estimates to the graph; 3) Filtering and temporal placement of gestures using text-gesture associations and timing profiles, respectively. The final output was an XML action script that could be used in a downstream animation system.

Extending this approach, Neff et al.~\cite{neff2008gesture} proposed a statistical system that learned gesture profiles but also added a character-specific animation lexicon. The system had two distinct phases. The pre-processing stage started with a video corpus of a character, hand-annotated in ANVIL \cite{kipp2001anvil}. The annotation process was similar to that of Kipp \cite{kipp2005gesture}, but with an additional English-speaking character. Given the annotated data, a gesture profile (a statistical model) and animation lexicon were created, where the latter consisted of hand orientation, torso posture and data for after-strokes (i.e. subsequent repeated hand movements after a prominent stroke), for each gesture lexeme. The full-automated generation phase had two distinct paths: 1) \textit{re-creation} that took in an annotated video as input and could re-create the gestures (observed in the video) in the animation system, useful for validating the annotations; 2) \textit{gesture generation} that could generate gesture from novel annotated text without the need for video input. Either path leveraged the character's gesture profile to generate a gesture script. The gesture script was used in the animation engine to generate the final animation either through a kinematic or dynamic simulation algorithm.

Bergman and Kopp proposed a different statistical approach for modeling the transformation of speech that describes objects into iconic gestures that resemble said objects \cite{bergmann2009increasing}. The proposed system generated coordinated speech and gesture by leveraging propositional and imagistic knowledge representations for content planning and concrete speech and gesture formulation. Their work involved dyadic conversations where one speaker gives spatial directions to another after exploring a virtual environment in VR. The study investigated what contextual factors are important in the formation of speech and gesture describing physical objects. As part of their framework, they developed a Bayesian network for gesture formulation. The Bayesian network defined a probability distribution over gesture properties such as indexing, placing, shaping, drawing and posturing. The probability distribution also took into account the idiosyncratic patterns for mapping visuospatial referents onto gesture morphology, i.e. the specific way an individual might index, shape or draw a gesture when describing a referent object. Gesture formulation resulted in fine-grained features including hand shape, wrist location, palm direction, extended finger direction, movement trajectory and direction. For the final animation, the framework leveraged the rule-based Articulated Communicator Engine (ACE) \cite{kopp2004synthesizing} to realize synchronized speech and gesture.

Bergman and Kopp closely followed up with a hybrid framework combining data-driven and model-based techniques to model iconic gestures using Bayesian \textit{decision} networks \cite{bergmann2009gnetic}. They used a similar corpus of dyadic interactions with spontaneous speech and gesture employed for direction giving and landmark description. The corpus was richly annotated with temporal segments, gesture morphology and references to objects for iconic gestures. Extending their earlier work that used Bayesian networks \cite{bergmann2009increasing}, they used Bayesian decision networks, supplemented by decision network nodes \cite{howard2005influence}. Bayesian decision networks enabled them to formulate gesture generation as a finite sequential decision problem by combining probabilistic and rule-based components. For example, the decision to include a certain gesture or the morphology of the gesture could be encoded as a decision node (activated by a rule) or chance node (activated by probability) with a specific probability distribution. To generate a gesture, the Bayesian network defined a probability distribution over gesture morphological features, based on object referent features, discourse context and the previously performed gesture.

Levine et al. proposed a hidden Markov model (HMM) to select the most suitable motion clip from a motion capture database, by using prosody-based features extracted from the original speech \cite{levine2009real}. The trained HMM used prosody cues to select the most appropriate gesture sub-units from the motion capture, ensuring that the chosen sub-units transition smoothly and are appropriate for the tone of the current utterance. However, directly associating prosody with gesture sub-units created a dependence on the quality and amount of training data, which made the system susceptible to overfitting. Levine et al.~\cite{levine2010gesture} improved upon the previous system by proposing ``gesture controllers'' that decoupled the kinematic properties of gestures (e.g. velocity, spatial extent) from their shape. Gesture controllers inferred gesture kinematics using a conditional random field (CRF) that analyzed the acoustic features in the input speech and learned a distribution over a set of hidden states. The hidden states encoded the latent structure of gesture kinematics without regard for the morphology of the gesture which reduced the number of false correlations and thus alleviated overfitting. Finally, a Markov Decision Process (MDP) took the hidden states and the distribution over them as input and used an optimal policy (learned via the reinforcement learning algorithm) to select the appropriate gesture clips.

Chiu et al.~\cite{chiu2011train} maintained the use of prosodic features to learn a probabilistic model for gesture generation. They restricted their study to learning gesture types that are associated with prosody, i.e. rhythmic movements (beats). The gesture generator was based on a modified Hierarchical Factored Conditional Restricted Boltzmann Machine (HFCRBM) \cite{taylor2009factored}. They first built a compact motion representation by training a conditional Restricted Boltzmann Machine (CRBM) using an unsupervised learning algorithm. Then the HFCRBM generator autoregressively took in the previous gesture representation and a sequence of audio features extracted from the original speech to generate the gesture representation for every time step, until the full motion sequence was completed. Finally, they smoothed discontinuities between frames by reducing the acceleration of wrist joints if they exceed a set threshold. However, their approach was restricted to rhythmic gestures and thus did not consider other commonly occurring gesture types such as iconic, pantomimes, deictic, emblematic and metaphoric.

Recently, Yang et al.~\cite{yang2020statistics} proposed a statistical motion-graph-based system that generated gestures and other body motions for dyadic conversations, that were well synchronized with novel audio clips. They constructed a motion graph that preserved the statistics of a database of recorded dyadic conversations. During generation, the graph was used to search for the most plausible motion sequence according to three constraints of audio-motion coordination in human conversations: 1) coordination to phonemic clause; 2) listener response; 3) partner's hesitation pause. The system adapted motion graphs, successfully employed in locomotion \cite{lee2002interactive,kovar2002motion,arikan2002interactive,treuille2007near}, for free-form conversation gestures with a lot more stylistic variation. Their conversational motion graph was significantly larger than that for locomotion due to the richness of conversational gestures. Given such a large graph, the system balanced search efficiency and style diversity by leveraging a stochastic greedy search algorithm to find a high-quality animation, well synchronized with the audio.

Statistical models provide more flexibility than rule-based systems and capture the non-determinism found in conversational gestures. In fact, a lot of the statistical principles (i.e. learning a probability distribution over the gesture data, through maximum likelihood estimation (MLE)) are still useful and relevant to most state-of-the-art methods, currently dominated by deep learning-based models. However statistical systems usually considered a limited number of independent variables based on painstakingly annotated gesture data. Deep learning-based models provide more flexibility through great representative capacity as well as making even fewer assumptions about the statistics of the underlying data. We describe this family of models next.

\subsection{Deep learning approaches} \label{subsec:deep-learning}

Deep learning-based generative models recently gained interest because of their ability to synthesize data from abstract representations of training datasets. They are increasingly prominent in character animation applications, including character control in games, and facial or gesture animation conditioned on speech and text in virtual agents. Such models typically make few assumptions about the underlying data distribution (except useful inductive biases), and learn their parameters to fit the data through gradient-based optimization of an objective function. 

\new{The use of deep-learning approaches has moved the field forward substantially in terms of perceived naturalness, but arguably represents a step backwards in terms of communicative efficacy with respect to previous methods, as illustrated in Figure \ref{fig:stefan_overview}. Similar to previous approaches, the main targets of deep learning systems have been human-likeness and appropriateness for speech audio and semantic content. The former is the degree to which the generated gesture motion visually resembles believable human behavior, while the latter is how suitable it is for a given speech audio, text input, or other contextual information. Early deep-learning systems ignored semantics, instead focusing on improving human-likeness \cite{alexanderson2020style, kucherenko2019analyzing, ferstl2020adversarial}. Later approaches have tried to incorporate semantics in order to generate meaningful gestures. The first attempts could generate only a handful of such gestures \cite{kucherenko2020gesticulator, yoon2020speech, ahuja2020no}. Although more recent work suggest that progress can \cite{kucherenko2022multimodal} and has been made \cite{ao2022rhythmic}, appropriateness remains a challenge. 
This can be seen from the GENEA Challenge (where GENEA stands for Generation and Evaluation of Non-verbal Behavior for Embodied Agents), which is a recurring large-scale comparison of gesture synthesis systems, whose most recent iteration \cite{yoon2022genea} found that
the human-likeness of motion can now reach the level of human motion capture, while appropriateness is still barely above chance.}

The proliferation of deep learning in conversational gesture generation has led to a large number of approaches that can be grouped based on the input modalities, i.e.\ audio, text, audio and text, or audio with other non-communicative modalities, and control parameters. We employ this taxonomy to organize our exposition and give a summary of the models and their respective categories in Table \ref{tab:dl-sota}. \new{We only include approaches that produce hand gestures and were published before the submission deadline for our review (October 21, 2022). In sections \ref{subsubsec:audio-input}, \ref{subsubsec:text-input}, and \ref{subsubsec:audio-text-input} we discuss generation approaches that use audio-only, text-only, and a combination of audio and text input, respectively. Section \ref{subsubsec:non-linguistic-input} focuses on approaches that use non-linguistic input, i.e. input other than speech audio or text. Finally, Section \ref{subsubsec:control-input} explores approaches that employ control input. The approaches within each modality section are presented in chronological order to reflect the evolution of the field.}

\begin{table*}[]
\centering
\tiny
\setlength\tabcolsep{1.25pt}
\begin{tabular}{|l|l|lllll|l|l|llll|l|}
\hline
\multicolumn{1}{|c|}{\multirow{2}{*}{\textbf{Model}}} & \multicolumn{1}{c|}{\multirow{2}{*}{\textbf{Dataset}}}                 & \multicolumn{5}{c|}{\textbf{Inputs}}                                                                                                                                                        & \multicolumn{1}{c|}{\multirow{2}{*}{\textbf{Training Objective}}} & \multicolumn{1}{c|}{\multirow{2}{*}{\textbf{Sequential}}} & \multicolumn{4}{c|}{\textbf{Output Representation}}                                                                                                                & \multicolumn{1}{c|}{\multirow{2}{*}{\textbf{Stochastic}}} \\ \cline{3-7} \cline{10-13}
\multicolumn{1}{|c|}{}                                & \multicolumn{1}{c|}{}                                                  & \multicolumn{1}{c|}{\textbf{Audio}} & \multicolumn{1}{c|}{\textbf{Text}} & \multicolumn{1}{c|}{\textbf{Pose}} & \multicolumn{1}{c|}{\textbf{Control}} & \multicolumn{1}{c|}{\textbf{Other}} & \multicolumn{1}{c|}{}                                             & \multicolumn{1}{c|}{}                                     & \multicolumn{1}{c|}{\textbf{Joint pos.}} & \multicolumn{1}{c|}{\textbf{Joint rot.}} & \multicolumn{1}{c|}{\textbf{Pre-rec.}} & \multicolumn{1}{c|}{\textbf{Other}} & \multicolumn{1}{c|}{}                                     \\ \hline
DCNF \cite{chiu2015predicting}                        & DIAC \cite{gratch2014distress}                                         & \multicolumn{1}{l|}{}               & \multicolumn{1}{l|}{\cmark}        & \multicolumn{1}{l|}{}              & \multicolumn{1}{l|}{}                 &                                     & MLE                                                               & \cmark                                                    & \multicolumn{1}{l|}{}                    & \multicolumn{1}{l|}{}                    & \multicolumn{1}{l|}{\cmark}            &                                     &                                                           \\ \hline
Hasegawa et al.~\cite{hasegawa2018evaluation}         & Gesture-Speech \cite{takeuchi2017creating}                             & \multicolumn{1}{l|}{\cmark}         & \multicolumn{1}{l|}{}              & \multicolumn{1}{l|}{}              & \multicolumn{1}{l|}{}                 &                                     & MSE                                                               & \cmark                                                    & \multicolumn{1}{l|}{\cmark}              & \multicolumn{1}{l|}{}                    & \multicolumn{1}{l|}{}                  &                                     &                                                           \\ \hline
Ishi et al.~\cite{ishi2018speech}                     & Ishi et al.~\cite{ishi2018speech}                                      & \multicolumn{1}{l|}{\cmark}         & \multicolumn{1}{l|}{\cmark}        & \multicolumn{1}{l|}{}              & \multicolumn{1}{l|}{}                 &                                     & MLE                                                               & \cmark                                                    & \multicolumn{1}{l|}{}                    & \multicolumn{1}{l|}{}                    & \multicolumn{1}{l|}{\cmark}            &                                     & \cmark                                                    \\ \hline
Kucherenko et al.~\cite{kucherenko2019analyzing}      & Gesture-Speech \cite{takeuchi2017creating}                             & \multicolumn{1}{l|}{\cmark}         & \multicolumn{1}{l|}{}              & \multicolumn{1}{l|}{}              & \multicolumn{1}{l|}{}                 &                                     & MSE                                                               & \cmark                                                    & \multicolumn{1}{l|}{\cmark}              & \multicolumn{1}{l|}{}                    & \multicolumn{1}{l|}{}                  &                                     &                                                           \\ \hline
Yoon et al.~\cite{yoon2019robots}                     & TED \cite{yoon2019robots}                                              & \multicolumn{1}{l|}{}               & \multicolumn{1}{l|}{\cmark}        & \multicolumn{1}{l|}{}              & \multicolumn{1}{l|}{}                 &                                     & MSE + MAE - Var                                                   & \cmark                                                    & \multicolumn{1}{l|}{\cmark}              & \multicolumn{1}{l|}{}                    & \multicolumn{1}{l|}{}                  &                                     &                                                           \\ \hline
DRAM \cite{ahuja2019react}                            & Ahuja et al.~\cite{ahuja2019react}                                     & \multicolumn{1}{l|}{\cmark}         & \multicolumn{1}{l|}{}              & \multicolumn{1}{l|}{\cmark}        & \multicolumn{1}{l|}{}                 &                                     & MSE                                                               & \cmark                                                    & \multicolumn{1}{l|}{}                    & \multicolumn{1}{l|}{}                    & \multicolumn{1}{l|}{\cmark}            &                                     &                                                           \\ \hline
Speech2Gesture \cite{ginosar2019learning}             & S2G \cite{ginosar2019learning}                                         & \multicolumn{1}{l|}{\cmark}         & \multicolumn{1}{l|}{}              & \multicolumn{1}{l|}{}              & \multicolumn{1}{l|}{}                 &                                     & MAE + Adv                                                         &                                                           & \multicolumn{1}{l|}{\cmark}              & \multicolumn{1}{l|}{}                    & \multicolumn{1}{l|}{}                  &                                     &                                                           \\ \hline
Ferstl et al.~\cite{ferstl2019multi}                  & Trinity I \cite{ferstl2018investigating}                               & \multicolumn{1}{l|}{\cmark}         & \multicolumn{1}{l|}{}              & \multicolumn{1}{l|}{}              & \multicolumn{1}{l|}{}                 &                                     & MSE + Adv                                                         & \cmark                                                    & \multicolumn{1}{l|}{\cmark}              & \multicolumn{1}{l|}{}                    & \multicolumn{1}{l|}{}                  &                                     & \cmark                                                    \\ \hline
CDBN \cite{sadoughi2019speech}                        & MSP-AVATAR \cite{sadoughi2015msp}                                      & \multicolumn{1}{l|}{\cmark}         & \multicolumn{1}{l|}{}              & \multicolumn{1}{l|}{}              & \multicolumn{1}{l|}{}                 & \cmark                              & EM                                                                & \cmark                                                    & \multicolumn{1}{l|}{}                    & \multicolumn{1}{l|}{}                    & \multicolumn{1}{l|}{}                  & \cmark                              & \cmark                                                    \\ \hline
Mix-StAGE \cite{ahuja2020style}                       & PATS \cite{ahuja2020style}                                             & \multicolumn{1}{l|}{\cmark}         & \multicolumn{1}{l|}{}              & \multicolumn{1}{l|}{}              & \multicolumn{1}{l|}{}                 &                                     & MAE + CCE + Adv                                                   &                                                           & \multicolumn{1}{l|}{\cmark}              & \multicolumn{1}{l|}{}                    & \multicolumn{1}{l|}{}                  &                                     & \cmark                                                    \\ \hline
Yoon et al.~\cite{yoon2020speech}                     & TED \cite{yoon2019robots}                                              & \multicolumn{1}{l|}{\cmark}         & \multicolumn{1}{l|}{\cmark}        & \multicolumn{1}{l|}{}              & \multicolumn{1}{l|}{}                 & \cmark                              & Huber + Adv + KL                                                  & \cmark                                                    & \multicolumn{1}{l|}{\cmark}              & \multicolumn{1}{l|}{}                    & \multicolumn{1}{l|}{}                  &                                     & \cmark                                                    \\ \hline
Bozkurt et al. \cite{bozkurt2020affective}            & Creative IT \cite{metallinou2016usc}                                   & \multicolumn{1}{l|}{\cmark}         & \multicolumn{1}{l|}{}              & \multicolumn{1}{l|}{}              & \multicolumn{1}{l|}{}                 & \cmark                              & MLE                                                               & \cmark                                                    & \multicolumn{1}{l|}{}                    & \multicolumn{1}{l|}{}                    & \multicolumn{1}{l|}{}                  & \cmark                              &                                                           \\ \hline
Gesticulator \cite{kucherenko2020gesticulator}        & Trinity I  \cite{ferstl2018investigating}                              & \multicolumn{1}{l|}{\cmark}         & \multicolumn{1}{l|}{\cmark}        & \multicolumn{1}{l|}{}              & \multicolumn{1}{l|}{}                 &                                     & MSE                                                               & \cmark                                                    & \multicolumn{1}{l|}{}                    & \multicolumn{1}{l|}{\cmark}              & \multicolumn{1}{l|}{}                  &                                     &                                                           \\ \hline
StyleGestures \cite{alexanderson2020style}            & Trinity I \cite{ferstl2018investigating}                               & \multicolumn{1}{l|}{\cmark}         & \multicolumn{1}{l|}{}              & \multicolumn{1}{l|}{}              & \multicolumn{1}{l|}{\cmark}           &                                     & MLE                                                               & \cmark                                                    & \multicolumn{1}{l|}{\cmark}              & \multicolumn{1}{l|}{}                    & \multicolumn{1}{l|}{}                  &                                     & \cmark                                                    \\ \hline
AiSLE \cite{ahuja2020no}                              & PATS \cite{ahuja2020style}                                             & \multicolumn{1}{l|}{\cmark}         & \multicolumn{1}{l|}{\cmark}        & \multicolumn{1}{l|}{}              & \multicolumn{1}{l|}{}                 &                                     & MAE + CCE + Adv                                                   &                                                           & \multicolumn{1}{l|}{\cmark}              & \multicolumn{1}{l|}{}                    & \multicolumn{1}{l|}{}                  &                                     & \cmark                                                    \\ \hline
Habibie et al.~\cite{habibie2021learning}             & S2G 3D \cite{ginosar2019learning,habibie2021learning}                  & \multicolumn{1}{l|}{\cmark}         & \multicolumn{1}{l|}{}              & \multicolumn{1}{l|}{}              & \multicolumn{1}{l|}{}                 &                                     & MSE + MAE + Adv                                                   &                                                           & \multicolumn{1}{l|}{\cmark}              & \multicolumn{1}{l|}{}                    & \multicolumn{1}{l|}{}                  &                                     &                                                           \\ \hline
Korzun et al.~\cite{korzun2021audio}                  & Kucherenko et al.~\cite{kucherenko2020genea}                           & \multicolumn{1}{l|}{\cmark}         & \multicolumn{1}{l|}{\cmark}        & \multicolumn{1}{l|}{}              & \multicolumn{1}{l|}{}                 &                                     & MSE                                                               & \cmark                                                    & \multicolumn{1}{l|}{}                    & \multicolumn{1}{l|}{\cmark}              & \multicolumn{1}{l|}{}                  &                                     &                                                           \\ \hline
Audio2Gestures \cite{li2021audio2gestures}            & Trinity I  \cite{ferstl2018investigating}                              & \multicolumn{1}{l|}{\cmark}         & \multicolumn{1}{l|}{}              & \multicolumn{1}{l|}{}              & \multicolumn{1}{l|}{}                 &                                     & MAE + GeoD + KL                                                   &                                                           & \multicolumn{1}{l|}{\cmark}              & \multicolumn{1}{l|}{\cmark}              & \multicolumn{1}{l|}{}                  &                                     & \cmark                                                    \\ \hline
Body2Hands \cite{ng2021body2hands}                    & S2G \cite{ginosar2019learning}                                         & \multicolumn{1}{l|}{}               & \multicolumn{1}{l|}{}              & \multicolumn{1}{l|}{\cmark}        & \multicolumn{1}{l|}{}                 & \cmark                              & MAE + Adv                                                         &                                                           & \multicolumn{1}{l|}{\cmark}              & \multicolumn{1}{l|}{}                    & \multicolumn{1}{l|}{}                  &                                     &                                                           \\ \hline
Lee et al.~\cite{lee2021crossmodal}                   & PATS \cite{ahuja2020style}                                             & \multicolumn{1}{l|}{\cmark}         & \multicolumn{1}{l|}{\cmark}        & \multicolumn{1}{l|}{}              & \multicolumn{1}{l|}{}                 &                                     & MAE + Adv + CC-NCE                                                & \cmark                                                    & \multicolumn{1}{l|}{\cmark}              & \multicolumn{1}{l|}{}                    & \multicolumn{1}{l|}{}                  &                                     &                                                           \\ \hline
Text2Gestures \cite{bhattacharya2021text2gestures}    & MPI-EBEDB \cite{volkova2014mpi}                                        & \multicolumn{1}{l|}{}               & \multicolumn{1}{l|}{\cmark}        & \multicolumn{1}{l|}{}              & \multicolumn{1}{l|}{}                 & \cmark                              & MSE                                                               & \cmark                                                    & \multicolumn{1}{l|}{}                    & \multicolumn{1}{l|}{\cmark}              & \multicolumn{1}{l|}{}                  &                                     &                                                           \\ \hline
ExpressGesture \cite{ferstl2021expressgesture}        & Trinity A + B  \cite{ferstl2018investigating,ferstl2021expressgesture} & \multicolumn{1}{l|}{\cmark}         & \multicolumn{1}{l|}{}              & \multicolumn{1}{l|}{}              & \multicolumn{1}{l|}{}                 &                                     & MSE                                                               & \cmark                                                    & \multicolumn{1}{l|}{}                    & \multicolumn{1}{l|}{}                    & \multicolumn{1}{l|}{\cmark}            &                                     &                                                           \\ \hline
CMCF \cite{saund2021cmcf}                             & S2G \cite{ginosar2019learning}                                         & \multicolumn{1}{l|}{\cmark}         & \multicolumn{1}{l|}{\cmark}        & \multicolumn{1}{l|}{}              & \multicolumn{1}{l|}{}                 &                                     & WCSS                                                              & \cmark                                                    & \multicolumn{1}{l|}{}                    & \multicolumn{1}{l|}{}                    & \multicolumn{1}{l|}{\cmark}            &                                     &                                                           \\ \hline
Qian et al.~\cite{qian2021speech}                     & S2G \cite{ginosar2019learning}                                         & \multicolumn{1}{l|}{\cmark}         & \multicolumn{1}{l|}{}              & \multicolumn{1}{l|}{}              & \multicolumn{1}{l|}{}                 &                                     & MAE + KL                                                          &                                                           & \multicolumn{1}{l|}{\cmark}              & \multicolumn{1}{l|}{}                    & \multicolumn{1}{l|}{}                  & \cmark                              &                                                           \\ \hline
Rebol et al.~\cite{rebol2021passing}                  & Rebol et al.~\cite{rebol2021passing}                                   & \multicolumn{1}{l|}{\cmark}         & \multicolumn{1}{l|}{}              & \multicolumn{1}{l|}{}              & \multicolumn{1}{l|}{}                 &                                     & MAE + Adversarial                                                 &                                                           & \multicolumn{1}{l|}{\cmark}              & \multicolumn{1}{l|}{}                    & \multicolumn{1}{l|}{}                  &                                     &                                                           \\ \hline
Flow-VAE ~\cite{taylor2021speech}                     & Taylor et al.~\cite{taylor2021speech}                                  & \multicolumn{1}{l|}{\cmark}         & \multicolumn{1}{l|}{}              & \multicolumn{1}{l|}{}              & \multicolumn{1}{l|}{}                 &                                     & NLL + MSE                                                         & \cmark                                                    & \multicolumn{1}{l|}{}                    & \multicolumn{1}{l|}{\cmark}              & \multicolumn{1}{l|}{}                  &                                     & \cmark                                                    \\ \hline
Wu et al.~\cite{wu2021probabilistic}                  & Gesture-Speech \cite{takeuchi2017creating}                             & \multicolumn{1}{l|}{\cmark}         & \multicolumn{1}{l|}{}              & \multicolumn{1}{l|}{\cmark}        & \multicolumn{1}{l|}{}                 & \cmark                              & Adv + WGAN-GP + Huber                                             & \cmark                                                    & \multicolumn{1}{l|}{}                    & \multicolumn{1}{l|}{\cmark}              & \multicolumn{1}{l|}{}                  &                                     & \cmark                                                    \\ \hline
Wu et al.~\cite{wu2021modeling}                       & Gesture-Speech \cite{takeuchi2017creating}                             & \multicolumn{1}{l|}{\cmark}         & \multicolumn{1}{l|}{}              & \multicolumn{1}{l|}{}              & \multicolumn{1}{l|}{}                 & \cmark                              & Adv                                                               & \cmark                                                    & \multicolumn{1}{l|}{\cmark}              & \multicolumn{1}{l|}{}                    & \multicolumn{1}{l|}{}                  &                                     & \cmark                                                    \\ \hline
Kucherenko et al.~\cite{kucherenko2022multimodal}     & SaGA++ \cite{kucherenko2022multimodal}                                 & \multicolumn{1}{l|}{\cmark}         & \multicolumn{1}{l|}{\cmark}        & \multicolumn{1}{l|}{}              & \multicolumn{1}{l|}{}                 &                                     & CCE                                                               &                                                           & \multicolumn{1}{l|}{}                    & \multicolumn{1}{l|}{}                    & \multicolumn{1}{l|}{}                  & \cmark                              &                                                           \\ \hline
Nguyen et al. \cite{nguyen2022context}                & JESTKOD \cite{bozkurt2017jestkod}                                      & \multicolumn{1}{l|}{\cmark}         & \multicolumn{1}{l|}{}              & \multicolumn{1}{l|}{\cmark}        & \multicolumn{1}{l|}{}                 &                                     & Adv                                                               & \cmark                                                    & \multicolumn{1}{l|}{}                    & \multicolumn{1}{l|}{}                    & \multicolumn{1}{l|}{\cmark}            &                                     &                                                           \\ \hline
GestureMaster \cite{zhou2022gesturemaster}            & TWH GENEA \cite{yoon2022genea}                                         & \multicolumn{1}{l|}{\cmark}         & \multicolumn{1}{l|}{\cmark}        & \multicolumn{1}{l|}{}              & \multicolumn{1}{l|}{}                 &                                     & L2 + Hamm + ETC                                                   & \cmark                                                    & \multicolumn{1}{l|}{}                    & \multicolumn{1}{l|}{}                    & \multicolumn{1}{l|}{\cmark}            & \cmark                              &                                                           \\ \hline
ZeroEGGS \cite{ghorbani2022exemplar}                  & Ghorbani et al.~\cite{ghorbani2022exemplar,ghorbani2022zeroeggs}       & \multicolumn{1}{l|}{\cmark}         & \multicolumn{1}{l|}{}              & \multicolumn{1}{l|}{}              & \multicolumn{1}{l|}{\cmark}           &                                     & MAE + KL                                                          & \cmark                                                    & \multicolumn{1}{l|}{\cmark}              & \multicolumn{1}{l|}{\cmark}              & \multicolumn{1}{l|}{}                  & \cmark                              & \cmark                                                    \\ \hline
ZS-MSTM \cite{fares2022zero}                          & PATS \cite{ahuja2020style}                                             & \multicolumn{1}{l|}{\cmark}         & \multicolumn{1}{l|}{\cmark}        & \multicolumn{1}{l|}{}              & \multicolumn{1}{l|}{}                 &                                     & MSE + Adv                                                         & \cmark                                                    & \multicolumn{1}{l|}{\cmark}              & \multicolumn{1}{l|}{}                    & \multicolumn{1}{l|}{}                  &                                     &                                                           \\ \hline
Habibie et al.~\cite{habibie2022motion}               & S2G 3D \cite{ginosar2019learning,habibie2021learning}                  & \multicolumn{1}{l|}{\cmark}         & \multicolumn{1}{l|}{}              & \multicolumn{1}{l|}{}              & \multicolumn{1}{l|}{\cmark}           &                                     & MAE + Adv + WGAN-GP                                               &                                                           & \multicolumn{1}{l|}{\cmark}              & \multicolumn{1}{l|}{}                    & \multicolumn{1}{l|}{}                  &                                     & \cmark                                                    \\ \hline
SEEG \cite{liang2022seeg}                             & TED \cite{yoon2019robots}                                              & \multicolumn{1}{l|}{\cmark}         & \multicolumn{1}{l|}{\cmark}        & \multicolumn{1}{l|}{}              & \multicolumn{1}{l|}{}                 &                                     & MAE + Adv + KL                                                    &                                                           & \multicolumn{1}{l|}{}                    & \multicolumn{1}{l|}{}                    & \multicolumn{1}{l|}{}                  &                                     &                                                           \\ \hline
Zhuang et al.~\cite{zhuang2022text}                   & Zhuang et al.~\cite{zhuang2022text}                                    & \multicolumn{1}{l|}{\cmark}         & \multicolumn{1}{l|}{\cmark}        & \multicolumn{1}{l|}{}              & \multicolumn{1}{l|}{}                 &                                     & MSE + SIMM + ETC                                                  &                                                           & \multicolumn{1}{l|}{}                    & \multicolumn{1}{l|}{}                    & \multicolumn{1}{l|}{\cmark}            &                                     &                                                           \\ \hline
Zhou et al.~\cite{zhou2022audio}                      & Personal Story ~\cite{zhou2022audio}, TED \cite{yoon2019robots}        & \multicolumn{1}{l|}{\cmark}         & \multicolumn{1}{l|}{}              & \multicolumn{1}{l|}{}              & \multicolumn{1}{l|}{}                 & \cmark                              & MAE                                                               & \cmark                                                    & \multicolumn{1}{l|}{}                    & \multicolumn{1}{l|}{}                    & \multicolumn{1}{l|}{}                  & \cmark                              &                                                           \\ \hline
Deichler et al.~\cite{deichler2022towards}            & Deichler et al.~\cite{deichler2022towards}                             & \multicolumn{1}{l|}{}               & \multicolumn{1}{l|}{}              & \multicolumn{1}{l|}{\cmark}        & \multicolumn{1}{l|}{}                 &                                     & IR + TR                                                           & \cmark                                                    & \multicolumn{1}{l|}{\cmark}              & \multicolumn{1}{l|}{}                    & \multicolumn{1}{l|}{}                  &                                     &                                                           \\ \hline
DiffGAN \cite{ahuja2022low}                           & PATS \cite{ahuja2020style}                                             & \multicolumn{1}{l|}{\cmark}         & \multicolumn{1}{l|}{\cmark}        & \multicolumn{1}{l|}{\cmark}        & \multicolumn{1}{l|}{}                 &                                     & MAE + MSE + Adv                                                   &                                                           & \multicolumn{1}{l|}{\cmark}              & \multicolumn{1}{l|}{}                    & \multicolumn{1}{l|}{}                  &                                     &                                                           \\ \hline
Rhythmic Gesticulator \cite{ao2022rhythmic}     & Trinity I \cite{ferstl2018investigating}, TED \cite{yoon2019robots}    & \multicolumn{1}{l|}{\cmark}         & \multicolumn{1}{l|}{\cmark}        & \multicolumn{1}{l|}{\cmark}        & \multicolumn{1}{l|}{\cmark}           &                                     & MSE + KL                                                          & \cmark                                                    & \multicolumn{1}{l|}{}                    & \multicolumn{1}{l|}{\cmark}              & \multicolumn{1}{l|}{}                  &                                     & \cmark                                                    \\ \hline
\end{tabular}
\caption{Summary of deep learning-based models in chronological order. The ``Sequential'' column indicates frame-by-frame generation otherwise frames are generated in parallel. The ``Stochastic'' column indicates varied gesture output for any input otherwise the output is the deterministic. See Table \ref{tab:obj-abbrv} for elaborations of the ``Training Objective'' abbreviations.}
\label{tab:dl-sota}
\end{table*}

\begin{table}[]
    \centering
    \begin{tabular}{|c|c|} 
        \hline
         \textbf{Objective} & \textbf{Full name} \\ 
         \hline
        Adv & Adversarial Loss \\
        CCE & Categorical Cross Entropy \\
        CC-NCE & Cross-modal Cluster Noise Contrastive Estimation \\
        ETC & Edge Transition Cost \\
        EM & Expectation Maximization \\
        GeoD & Geodesic Distance \\
        WGAN-GP & Wasserstein-GAN Gradient Penalty \\
        Hamm & Hamming Distance \\
        Huber & Huber Loss \\
        IR & Imitation Reward \\
        KL & Kullback–Leibler Divergence \\
        L2 & L2 Distance \\
        MAE & Mean Absolute Error \\
        MLE & Maximum Likelihood Estimation \\
        MSE & Mean Squared Error \\
        NLL & Negative Log-likelihood \\
        SIMM & Structural Similarity Index Measure \\
        TR & Task Reward \\
        Var & Variance \\
        WCSS & Within-cluster Sum of Squares \\
        \hline
    \end{tabular}
    \caption{Training objective abbreviations in Table \ref{tab:dl-sota}}
    \label{tab:obj-abbrv}
\end{table}


\subsubsection{Audio input} \label{subsubsec:audio-input}

Hasegawa et al.~\cite{hasegawa2018evaluation} proposed an autoregressive approach to generate gesture from audio utterances using a bi-directional LSTM \cite{hochreiter1997long}. The bi-directional LSTM learned audio-gesture relationships with both backward and forward consistencies over a long period of time. The model was trained with a then novel audio-gesture dataset, collected using a headset and marker-based motion capture \cite{takeuchi2017creating}. The model predicted a full skeletal human pose from the utterance features input at every LSTM timestep. Temporal filtering was then used to smooth out discontinuities in the generated pose sequences.

Kucherenko et al.~\cite{kucherenko2019analyzing} extended the work of Hasegawa et al.~\cite{hasegawa2018evaluation}, removing the need for temporal smoothing through representation learning of an autoencoder. The proposed model transformed audio input into a gesture sequence in the form of 3D
joint coordinates. They achieved this by (i) learning a lower dimensional representation of human motion using a denoising autoencoder consisting of a motion encoder (called MotionE) and a motion decoder (called MotionD) and (ii) training a novel speech encoder (called SpeechE) to transform speech to the corresponding motion representation with reduced dimensionality. During inference, the SpeechE predicted the motion representations, based on a given speech signal, and the MotionD decoded the motion representations into gesture sequences. However, their approach was deterministic and thus unable to capture the commonly observed phenomena where a person gesticulates differently at different points of the same utterance.

Deterministic generative approaches usually learn their parameters using a regression objective, e.g. L1 (Mean Absolute Error) or L2 (Mean Squared Error). Optimizing with either of those objectives typically forces the model toward learning to generate the mean representation of the data, producing averaged motion for different inputs, and resulting in undesirable results; usually called regression to the mean. Several approaches avoided this by incorporating probabilistic components into their objectives. Probabilistic components can increase the range of gesture motion in multiple ways, namely: (i) greater range of motion for different inputs, or (ii) stochastic motion for the same input. The most prominent are implicit log-likelihood evaluation via adversarial learning with Generative Adversarial Networks (GAN) \cite{goodfellow2014generative}, explicit log-likelihood evaluation via variational inference with Variational Autoencoders (VAE)\cite{kingma2013auto}, and exact log-likelihood evaluation via invertible transformations with Normalizing Flows \cite{kobyzev2020normalizing, papamakarios2021normalizing}.
 
GANs aim to do implicit density estimation of the underlying distribution through the interplay of a generator that tries to produce samples that are representative of the data, and a discriminator that strengthens the generator by classifying samples as real (from the distribution) or fake (not from the distribution). Multiple gesture generation approaches added an adversarial objective as a term in a composite loss function, which increased the range of gesture motion although still deterministic for a given audio input \cite{sadoughi2018novel,ginosar2019learning,ferstl2020adversarial,yoon2020speech,ahuja2020style,rebol2021passing,wu2021probabilistic,wu2021modeling,zabala2022modeling,habibie2022motion}. We discuss some notable examples below.

Ferstl et al.~\cite{ferstl2019multi,ferstl2020adversarial} added multiple adversarial objectives to a recurrent neural network, known to be susceptible to regression to the mean for long sequences. The adversaries accounted for gesture phase structure, motion realism, displacement and diversity of the minibatch. Ahuja et al.~\cite{ahuja2020style} learned to implicitly estimate a mixture of densities, each representing a speaker to enable the transfer of one speaker's style onto the speech input of another. The mixture of densities was estimated by instantiating a generator (per speaker) that is responsible for generating gestures that are representative of that speaker's underlying gesture distribution. All the generators were trained in unison using the adversarial objective. Most recently, Habibie et al.~\cite{habibie2022motion} used an adversarial objective in the form of a conditional GAN to refine gesture clips that were selected using a k-Nearest Neighbour (kNN) search that is conditioned on speech and a control signal.

Normalizing flows learn to describe highly complex distributions by applying invertible sub-transformations to a simple initial distribution, where invertibility allows one to optimize the exact likelihood of the deep generative model via gradient descent \cite{kobyzev2020normalizing, papamakarios2021normalizing}, unlike in GANs or VAEs. In the context of gesture generation, Alexanderson et al.~\cite{alexanderson2020style} extended a normalizing flow-based model for locomotion \cite{henter2020moglow} to apply to speech audio-driven gesture synthesis with style control.
Their normalizing flows learned invertible transformations from simple Gaussian distributions to the distribution of upper- or full-body motion capture data. These transformations were conditioned on speech acoustics and, optionally, arbitrary style parameters.
Specific style parameters considered included properties such as the average hand height, gesture speed, and gesture radius in a 4-second interval.
The resulting model produced gestures that scored among the best in terms of naturalness and appropriateness in the GENEA Challenge 2020 \cite{kucherenko2020genea}.

VAEs aim to do explicit density estimation of the underlying data by optimizing a combination of a reconstruction loss for an autoencoder (usually an L1 or L2 loss) and the Kullback-Leibler (KL) divergence for distribution matching between a prior distribution and approximate posterior distribution of the data. The prior distribution is usually instantiated as a Gaussian for simple parameterization. The learned stochastic variables can then be sampled and decoded into diverse outputs. Recently, Li et al.~\cite{li2021audio2gestures} used a conditional VAE to translate speech into diverse gestures. They explicitly modeled a one-to-many speech-to-gesture mapping by splitting the cross-modal latent code into a shared code (audio + gesture) and motion-specific code (gesture only). The shared code modeled the correlation between speech and gesture e.g. synchronized speech and rhythmic gestures. The motion-specific code attempted to capture the diversity in gesticulation, independent of audio information.

In a similar vein, Ghorbani et al.~\cite{ghorbani2022exemplar,ghorbani2022zeroeggs}, used a VAE-based framework for style controllable co-speech gesture generation conditioned by a zero-shot motion example i.e., an instance of a motion style unseen during training. Given an audio input and a motion example, they generated an encoding of the audio and a style embedding from the motion, and the two latent codes were used to guide the generation of stylized gestures. The variational nature of the style embedding enabled them to easily modify style through latent space manipulation or blending and scaling of style embeddings. Moreover, the probabilistic nature of the model enabled the generation of varied gestures for any audio and exemplar motion input. The resulting model performed favorably against state-of-the-art probabilistic techniques \cite{alexanderson2020style} in terms of naturalness of motion, appropriateness for speech, and style portrayal.

Taylor et al. \cite{taylor2021speech} adapted the conditional Flow-VAE framework \cite{bhattacharyya2019conditional}, combining the advantages of the VAE and normalizing flow architectures,to generate spontaneous gesture movement for speaker and listener roles in a dyadic interaction. They used the Flow-VAE framework for modeling expressive gesture because of its ability to improve generative capacity of the VAE by estimating the latent space with a highly complex distribution using a normalizing flow, instead of the standard Gaussian. Their autoregressive framework was trained on a set of previously generated gestures, an audio input window and the dyadic role, i.e. speaker or listener, as input. The preceding gestures were encoded into a latent variable then transformed into a complex distribution using a normalizing flow, conditioned by the audio window and role in the dyad. Their decoder then generated the next gesture based on the latent variable sampled from the complex distribution. The resulting model could generate expressive co-verbal gestures in a dyadic setting based.

Hybrid systems that combine deep learning and database matching components can also help tackle the regression to the mean problem \cite{kucherenko2022multimodal, zhou2022gesturemaster, ferstl2021expressgesture, habibie2022motion}. Indeed, this approach has been used effectively in motion synthesis problems, e.g. game animation where high fidelity motion is crucial \cite{holden2020learned}. In the context of conversational gesture, the intuition is that modeling the association between high dimension audio input and gestures, represented by exact joint positions or angles, using standard regression objectives (L1 or L2 loss) discourages the model from producing otherwise plausible gestures that do not exactly match the ground truth, thus greatly reducing the variety of generated gestures. Alternatively, the audio-gesture association can be modeled by predicting higher-level parameters for gesture motion. Ferstl et al.~\cite{ferstl2021expressgesture} realized this idea by learning to map audio to gesture via higher-level expressive parameters, specifically gesture velocity, acceleration, size, arm swivel angle, and extent of hand opening. First they pre-trained a model to associate audio prosodic features to the expressive parameters. Then they predicted gesture timing by extracting pitch peaks in the audio signal. At inference time, the prosodic features were used to estimate the expressive parameters that were in turn used to search for a matching gesture in the database, and the pitch peaks were used to temporally position the matching gesture. Finally, synthetic preparation and retraction phases were added to connect the gestures in the sequence.

Another interesting approach for preserving gesture form is through audio-based search in a video gesture database where the gestures are representated by video frames. Zhou et al.~\cite{zhou2022audio} explored this idea in a gesture reenactment task by generating a gesture video for an unseen audio input, using gesture frames from a reference video. They first encoded the reference gesture video as a ``video motion graph'' - a directed graph where each node represented a video frame and corresponding audio features, and the edges represented transitions. The graph encoded how the reference video can be split and re-assembled in difference graph paths. In order to increase graph connectivity, i.e, diversity of plausible path, they added synthetic edges based on a frame pose similarity threshold computed using the SMPL pose parameters \cite{loper2015smpl}. Given unseen audio input as a guide, they traversed the graph using  a beam search algorithm \cite{rubin1977locus} to find the most optimal path or order of gesture frames that best matches the speech audio. For graph paths that contain temporally disjoint frames, they trained pose-aware video blending network to synthesize smooth transitions between the frames.

\subsubsection{Text input} \label{subsubsec:text-input}

Approaches that used audio as the primary modality produced well-timed hand movements that tend to be highly associated with acoustics, largely corresponding to beat gestures. However, the lack of text transcript means they were not informed by the structure and context inherent in the text, for example, semantic meaning and punctuation. Such structure can help produce more meaningful and communicative gestures. Therefore, next, we describe some approaches that used text as the primary input modality.
 
Ishi et al.~\cite{ishi2018speech} proposed a text-based gesture generation approach for controlling a humanoid robot. They modeled the text-to-gesture motion translation by associating words to concepts, concepts to gesture categories (i.e. iconic, metaphoric, deictic, beat, emblem and adapter), and gesture categories to gesture motions. Further, they estimated conditional probabilities to model the association between word concepts and gesture categories, and between gesture categories and gesture motion clusters that were pre-computed with the k-means clustering algorithm.

Yoon et al.~\cite{yoon2019robots} proposed an encoder-decoder approach that transformed speech text, from a dataset based on TED talks, into a sequence of gestures. They created the TED video dataset by picking video segments with the speaker’s upper body and hands. Then they performed pose estimation using OpenPose \cite{cao2019OpenPose} and removed segments containing noisy or no estimations. Speech text was converted into a sequence of 300-dimensional word vectors using pre-trained GloVe embeddings \cite{pennington2014glove}. Similarly, poses estimations were converted to 10-dimensional vectors using principal component analysis (PCA). Therefore, co-speech gesture generation became a sequence-to-sequence translation problem from word embeddings to human poses encodings. The encoder part of the network was a bi-directional GRU \cite{cho2014learning} taking in speech text one word (vector) at a time and capturing bi-directional context. The last hidden state of the encoder was passed into the decoder, also a bi-directional GRU. The decoder also took previous pose estimations to condition the prediction of the next pose, in addition to using soft-attention \cite{bahdanau2014neural} to focus on specific words when predicting the next pose. Finally, the generated 2D poses were mapped to 3D and executed on the NAO humanoid robot.

Recently, Bhattacharya et al.~\cite{bhattacharya2021text2gestures} used text transcripts to produce expressive emotive gestures for virtual agents in narration and conversation settings, using MPI-EBEDB, a dataset of actors performing multiple emotion categories (amusement, anger, disgust, fear, joy, neutral, pride, relief, sadness, shame, surprise) \cite{volkova2014mpi}. Their approach consisted of Transformer-based encoders and decoders \cite{vaswani2017attention}, where the encoder took in the text transcript sentences (encoded as GloVe embeddings \cite{pennington2014glove}) to produce an encoding which was concatenated with the agent attributes such as narration/conversation, intended emotion, gender and handedness. The previous pose's encoded concatenation and 3D joint positions were passed as input to the Transformer decoder to generate the next pose's joint positions. The process was repeated in a recurrent manner until the full pose sequence was generated.

\subsubsection{Audio and text input} \label{subsubsec:audio-text-input}
An interesting trade-off exists between audio-based and text-based gesture generation systems. audio-based generators have access to intonation and prosody which helps generate rhythmic or kinematic gestures (e.g. beats) but lack semantic context. Conversely, text-based generators have access to semantic context which helps generate meaning-carrying gestures (e.g. iconic or metaphoric), but lack intonation and prosodic information. Therefore, combining the audio and text modalities enables a gesture generator to learn to produce semantically relevant and rhythmic co-speech gestures.

\new{Although generating meaning-carrying gestures using audio only is theoretically possible, it is unlikely since prosody is suitable for kinematics, but not sufficient to infer shape which is associated with meaning \cite{levine2012continuous}. As far as we know, meaningful gestures from speech audio  alone have not been empirically demonstrated. Instead, combining audio with text appears to be the most promising approach to generating meaningful gestures to date. We, therefore, focus on approaches that combine these two modalities for generating meaning-carrying, communicative gestures.}

Chiu et al.~\cite{chiu2015predicting} proposed an approach that combined the text and prosody of the speech to generate co-verbal gestures. Their model, called the Deep Conditional Neural Field (DCNF), was a combination of a fully-connected network, for representation learning, and a Conditional Random Field (CRF), for temporal modeling. For the gesture prediction task, the model took in a text transcript, part-of-speech tags and prosody features as input, and predicted a sequence of gesture signs which were a set of predefined hand motions.

Leveraging the representation power of deep learning models for multimodal input (i.e. audio and text) for co-speech gesture generation was the next logical step. In fact, three groups of researchers independently proposed the first deep-learning-based gesture generators that used both audio and text to generate continuous gestures, namely Yoon et al.~\cite{yoon2020speech}, Ahuja et al.~\cite{ahuja2020no} and Kucherenko et al.~\cite{kucherenko2020gesticulator}. We discuss their pioneering work combining audio and text next, followed by subsequent efforts in the area.

Yoon et al.~\cite{yoon2020speech} proposed a gesture generation approach that combines the tri-modal context of speech, text and speaker identity to produce gestures that were human-like, and matched the content and rhythm of the speech. The model processed input speech and text with a speech and text encoder, respectively. The speaker identity was used to sample the intended speaker from a learned style embedding space. Together the three features (i.e.\ speech encoding, text encoding, and style) were passed to a gesture generator to produce the sequence of poses.
Closely related was Liang et al.~\cite{liang2022seeg} whose framework utilized audio and text information in order to generate meaningful gestures by disentangling semantic and beat gestures. Their system consisted of two encoders, one that took in audio and text to encode semantics, and another that took in audio volume and beat to encode non-semantic information. The encoded information from both encoders ensured the disentanglement of semantic and beat gestures, while decoder took this information and was trained to encourage generation of meaningful semantic gestures.

Ahuja et al.~\cite{ahuja2020no} identified two key challenges in an attempt to learn the latent relationship between speech and co-speech gestures. First, the underlying distributions for text and gesture are inherently skewed and therefore necessitated the need to learn their respective long tails, accounting for rarely occurring text or gestures. Second, gesture predictions are made at the sub-word level, which necessitated the need to learn the relationship between language and acoustic cues that may give rise to, or be accompanied by, a particular gesticulation. So motivated, they proposed the Adversarial Importance Sampled Learning (AISLe) framework, that combined adversarial learning with importance sampling to balance precision and coverage. The model took in speech and text transcripts and performed encoding and alignment between sub-words and acoustics, using a multi-scale Transformer \cite{vaswani2017attention}. The resulting alignment was passed to the model's generator to predict the pose sequence and an adversarial discriminator was used to determine if the pose was real or fake. For optimizing the adversarial objective, the AISLe framework scaled the loss function such that rarely occurring gesture samples, the long tail of the distribution, were weighted more than those that are more likely to occur.

Kucherenko et al.~\cite{kucherenko2020gesticulator} proposed an autogressive generative model that combined speech acoustics and semantics to produce arbitrary acoustically-linked or semantically-linked gestures. The key insight of their approach was to envision a gesticulation system that encompasses so called ``representational'' gesture types (i.e.\ iconic, metaphoric and deictic) that convey semantics, and beats that are synchronized with acoustics. Their approach took a concatenation of semantic features that were extracted using BERT \cite{devlin2018bert} and acoustic features represented as log-power mel-spectograms as input into an encoder. Then they integrated past and future context for each gesture pose frame via a sliding window operation over the encoded speech features. The model generated each pose autoregressively where each was conditioned on the information of three preceding frames to ensure motion continuity. Their extensive evaluation indicated that autoregression for continuous motion and combining audio and text had the most significant positive impact on the quality of the generated gesticulations.

\new{Equally inspired by autoregressive generative models, Korzun et al.~\cite{korzun2021audio,vladislav_korzun_2020_4088609} reimplemented the text-only recurrent framework by \cite{yoon2019robots} to accommodate both text and audio input.} The proposed model was a combination of a recurrent context encoder\new{, inspired by \cite{kucherenko2019analyzing}}, that generated hidden states for 3-second audio and text context windows and a \new{recurrent} encoder-decoder that took in the concatenated results of the context encoder and used an attention mechanism to condition the generation of the final gesture motion. Similar to Yoon et al.~\cite{yoon2019robots}, they trained the model using the continuity and variance objectives to ensure fluid and natural-looking gestures. The resulting model produced gestures that were deemed natural and appropriate as part of the GENEA Challenge 2020 \cite{kucherenko2020genea}.

\new{Designing generation systems that produce meaningful gestures is one of the major goals in non-verbal behavior research. Spurred on by this question,} Kucherenko et al.~\cite{kucherenko2022multimodal} investigated whether contemporary deep learning-based systems could predict gesture properties, namely phase, type and semantic features, as a way to determine if such systems can consistently generate gestures that convey meaning. Their model used both audio and text for predicting gesture properties, through two distinct components
\new{that} predicted the probability for gesticulation
and probabilities for the \new{aforementioned} set of gesture properties. They conducted their experiments on a direction-giving dataset with a high number of representational gestures \cite{lucking2013data}. Their experiments showed that gesture properties related to meaning such as semantic properties and gesture type could be predicted from text features (encoded as FastText embeddings \cite{bojanowski2017enriching}), but not from prosodic audio features. Conversely, they found that rhythm-related gesture properties (e.g. phase) could be better predicted from audio features.

\new{In order to mimic the communicative intent of co-speech gestures, it is crucial to understand and model the complex relationship between speech acoustics, text, and hand movements. An interesting approach is to group gestures with distinct movement properties in order to find emergent rhetorical categories.} Saund et al.~\cite{saund2021cmcf} investigated \new{this approach} by modeling the rhetorical, semantic, affective and acoustic relationships between gestures and co-occurring speech audio and text, for a hypothetical gesture generation system. They first used k-means clustering to cluster speech-gesture pairs into functional domain clusterings (i.e. rhetorical, affective and semantic) based on functional tags generated from third-party natural language parsers. The speech-gesture pairs were refined into sub-clusters based on gesture motion. Therefore each speech-gesture pair belonged to at least one sub-cluster (based on motion), within one functional cluster (based on its assigned functional tags). At run-time, a hypothesized virtual agent would leverage the same pre-trained parsers and clusters to analyze an input speech and text transcription, select a functional cluster and from that a motion sub-cluster. The agent could then either choose an appropriate gesture from a pre-recorded library or the centroid gesture in the motion sub-cluster.

\new{Motion graphs, commonplace in conventional animation systems (e.g.\ \cite{kovar2002motion,arikan2002interactive,lee2002interactive}), can be effective at producing realistic non-verbal behavior because they rely on databases of high-quality motion capture or RGB video. As we discussed before, they were effectively employed for audio-driven gesture reenactment using video-based motion graphs \cite{zhou2022audio}.} Zhou et al.\ \new{continued this trend for audio and text by adapting a} motion-graph-based music-to-dance system \cite{chen2021choreomaster} for co-speech gesture generation ~\cite{zhou2022gesturemaster}. They first built a database of audio, text and gesture clips from 3-tuples of (audio, text transcript, gesture), using a splitting algorithm. For each audio clip, they generated a style signature using StyleGestures \cite{alexanderson2020style}, and a rhythm signature using a binary encoding scheme that denotes the presence of words by leveraging the word-level timing information in the text transcript. For the corresponding gesture motion, they generated a style signature, parameterized by the same attributes as StyleGestures \cite{alexanderson2020style} (e.g.\ wrist speed, radius and height), and a rhythm signature using a similar binary scheme that denoted the presence of pausing, sharp turning or a stroke phase of the gesture. During synthesis, they computed the rhythm and style signatures for input audio and text and used a graph-optimization algorithm to find gesture clips that closely matched the generated style and rhythm in terms of Hamming distance, and minimized the motion transition in the graph. This model performed on par or better than motion capture data in terms of Naturalness in the GENEA Challenge 2022 \cite{yoon2022genea}.

Style transfer is a widely adopted optimization technique in deep learning for blending visual content and style, e.g.\ given a content image and a reference image that specifies the style, adjust the content image to match the style \cite{gatys2016image}. In the context of co-speech gesture generation, it might be desirable to transfer the speaking style of one speaker to the predicted gestures of another. Ahuja et al.~\cite{ahuja2020style} learned unique style embeddings for multiple speakers that enabled either generation of gestures consistent with the original speaker in the audio input, or style transfer by combining the audio input of one speaker with the style embedding of a different speaker. Although they proposed the PATS data where multiple modalities such as audio, gesture pose and text have style and content, they focused on gesture pose style to learn \textit{unimodal} speaker-specific style embeddings. Fares et al.~\cite{fares2022zero} leveraged the multiple modalities in the PATS dataset to learn \textit{multimodal} style embeddings based on audio, text and gesture pose input. Their framework consisted of a speaker-style encoder that used speaker audio, text and gesture pose to learn a multimodal style embedding, and a sequence-to-sequence decoder that generated gestures based on audio and text, and conditioned on the desired speaker's style embedding. Furthermore, unlike the work of Ahuja et al.~\cite{ahuja2020style} that required the entire speaker's gesture data to learn the speaker's style embedding, their trained speaker-style encoder could generate style embeddings in a zero-shot manner i.e., for speaker styles not seen in the training set.

\new{A key tenet of semantically meaningful gestures is that they are appropriate for the given utterance. To achieve this, there needs to be a greater emphasis on generating precise gestures using audio and text as grounding (i.e. the appropriateness of the gesture to the utterance), versus generating diverse gestures. }Lee et al.~\cite{lee2021crossmodal} \new{investigated this approach and} made an interesting observation about human gesticulation, that multiple semantically different utterances are often accompanied by the same gesture. They thus proposed a contrastive-learning framework that constrained the mapping of semantically different utterances to a smaller subset of relevant high-quality gestures. They introduced a novel contrastive learning objective that preserved similarities and dissimilarities of gestures in the latent representation. The objective ensured that latent language representations of two semantically different utterances were close together if they were accompanied by the same gesture. They first clustered gestures based on similarity or dissimilarity, then created positive (similar gesture poses) and negative (dissimilar gesture poses) required for the standard contrastive learning objective. Finally, they learned gesture-aware embeddings via a contrastive and adversarial objective. The resulting embedding space was used to generate gestures that were semantically relevant and closer to the ground truth.

When designing 3D avatars, it may be desirable to have a holistic animation system that includes facial and full-body movement. Combining audio and text modalities can be effective at achieving this goal because of their rhythmic and semantic properties. Zhuang et al.~\cite{zhuang2022text} investigated this approach by proposing a hybrid system consisting of Transformer-based encoder and decoder modules, and motion-graph retrieval module to generate facial motion and full-body motion that included gestures. Their encoder used both audio and text, in the form of phoneme labels and Mel Frequency Cepstral Coefficients (MFCC) and Mel Filter Bank (MFB) features, to generate 3D facial parameters for synchronous lip movement. Simultaneously, the decoder used speech features, previous expression motion and semantic tags to generate 3D facial parameter for expression. The motion-graph retrieval sub-system used speech audio and text to find the most appropriate body motion segments, including gesture, that correspond to the text semantics and rhythm in the audio. Finally the facial and body motion were used to drive a skinned polygonal model.

\subsubsection{Non-linguistic modalities} \label{subsubsec:non-linguistic-input}
Several deep learning-based systems complemented input audio or text with additional information that could reasonably be deemed relevant to co-speech gestures. This included speech context, speaker style, discourse or an interlocutor's movements. Sadoughi and Busso \cite{sadoughi2019speech} proposed a system that bridges rule-based and learning-based techniques in order to select gestures that are communicative and well synchronized with speech. They proposed a Dynamic Bayesian Network (DBN) which took in speech and two constraints to condition the generation. The constraints were: 1) discourse function, which restricts the model to behaviors that are characteristic of that discourse class (e.g. questions); 2) prototypical behaviors, which restricted the model to certain target gesticulations (e.g. head nods). Given constraints on prototypical behaviors, the approach could be embedded in a rule-based system as a behavior realizer creating head and hand trajectories that are temporally synchronized with speech.

In a dyadic conversation between interlocutors, there can be a lot of spontaneous non-verbal behavior that is influenced by the nature and tone of the interaction. Leveraging the co-adaptation of non-verbal behavior between interlocutors present in human-to-human interactions, cf. \cite{bergmann2012gestural,cienki2014multimodal,oben2016explaining}, can enable virtual agents to be naturally conversational and collaborative. Ahuja et al.~\cite{ahuja2019react} proposed the Dyadic Residual-Attention Model (DRAM), a framework that could interactively generate an avatar's gesticulation conditioned on its speech and also the speech and gesticulation of a human interlocutor in a telepresence setting. In order to generate natural behavior, the avatar had to consider its own speech as well as the speech and gesticulation of the human. The DRAM model generated natural dyadic behavior by taking in the speech and pose history of the avatar as well as the speech and pose history of the human to adapt the avatar's gesticulation accordingly.

The idea of conditioning the motion of a deep-learning-based agent on interlocutor speech and motion has subsequently been used in several other works. Jonell et al.~\cite{jonell2020letsfaceit} used a model based on normalizing flows for generating head motion and facial expression, while Nguyen and Celiktutan \cite{nguyen2022context} used conditional adversarial learning to drive full-body skeletons. Both of these works found that statistically significant improvements in generated behaviors were achieved by being interlocutor-aware.

A similarly interesting dyadic scenario is human-robot interaction where one of the interlocutors is a social robot. In this case, the robot must exhibit natural non-verbal behavior in order to be engaging and interesting. Therefore, it is desirable for the robot to mimic human non-verbal motion with gestures that are natural and communicative. Deichler et al.~\cite{deichler2022towards} investigated this idea by proposing a combination of a data-driven and physically-based reinforcement learning (RL) framework to generate pointing gestures learned from motion capture data. Given a diverse motion capture dataset of pointing gestures and corresponding targets, they trained RL control policies adapted from \cite{peng2018deepmimic,peng2021amp} to imitate human-like pointing motion while maximizing the reward based on pointing precision.

Automatic synthesis and animation of gestures that accompany affective verbal communication can endow virtual agents with emotional impetus. Bozkurt et al.~\cite{bozkurt2020affective} directly mapped emotional cues in speech prosody into affect-expressive gestures. They investigated the use of three continuous affect attributes (i.e. activation, valence and dominance) for the speech-driven synthesis of affective gesticulation. They proposed a statistical model based on hidden semi-Markov models (HSMM) where states were gestures, and observations were speech prosody and continuous affect attributes. They first estimated the affective state from speech prosody and then used the state and speech prosody to predict gesture clusters. The gesture segments were animated using a unit selection algorithm \cite{bozkurt2015affect}, and discontinuities were smoothed using an exponential smoothing function. Finally, the smoothed sequence was animated in Autodesk MotionBuilder.

\new{Text encodes important semantic information, potentially useful for conveying meaningful emotion through gesture, although it encodes fewer cues about emotional state compared to audio e.g., intonation and speech pauses. An interesting approach is to combine text with an intended emotion for affective gesture generation.} Bhattacharya et al.~\cite{bhattacharya2021text2gestures} \new{pursued this approach by } combining text transcripts \new{associated with narrative or conversational acting and emotion labels}, to produce expressive emotive gestures for virtual agents. The emotions represented were amusement, anger, disgust, fear, joy, neutral, pride, relief, sadness, shame and surprise\cite{volkova2014mpi}. Their approach consisted of a Transformer \cite{vaswani2017attention} encoder and decoder, where the encoder took in the text transcript sentences, \new{intended} emotional state and agent attributes (e.g. narration/conversation, intended emotion, gender, handedness). The previous pose's encoded concatenation and 3D joint positions were passed as input to the Transformer decoder to generate the next pose's joint positions. The process was repeated in a recurrent manner until the full affective pose sequence was generated.

A speaker's identity or style can affect how they gesticulate, as some speakers gesture a lot while others rarely do. Moreover, they may also prefer particular gesture forms, and use different hands or gesture sizes. Modeling such variation in non-verbal behaviour can help make virtual agents seem unique and have a personality. To this end, Yoon et al \cite{yoon2020speech} used speaker identity to guide gesture generation that matched the speaker's style. Their adversarial approach combined the tri-modal context of audio, text and speaker identity to produce gestures that were human-like, and matched the content and rhythm of the speech. The model processed input audio and text with an audio and text encoder, respectively. The speaker identity was used to sample the intended speaker from a learned style embedding space. Together the three features (i.e. audio encoding, text encoding, and style) were passed to a gesture generator to produce the sequence of poses. Similarly, Ahuja et al.~\cite{ahuja2020style} learned a mixture of adversarial generators, representing diverse gesticulation styles of speakers from talk-show hosts, lecturers and televangelists. Learning speaker-specific generators enabled one speaker's style to be aligned with, or transferred to, the audio of another speaker.

Developing robust deep-learning-based gesture generators requires large amounts of diverse gesture data from real world scenarios, captured either via motion capture or pose estimation from videos. However, capturing or estimating hand gestures is very challenging because of the intricate finger motion, relatively small size of hands with respect to the whole body and frequent self-occlusions \cite{han2018online,lee2019talking,jorg2020virtual}. In contrast, capturing body motion (up to and including the arms) is less error prone because the joints are further apart and the articulations are relatively simpler. Therefore, the ``upper body'' motion as a modality can be an informative prior for generating conversational hand gestures. Ng et al.~\cite{ng2021body2hands} investigated this idea while making the observation that body motion is highly correlated with hand gestures. Their proposed approach took in 3D upper-body motion (up to the wrist) and predicted 3D hand poses. In addition to upper-body motion, the model could take in 2D images of hands and produce the corresponding 3D hand pose estimations. Similar to \cite{ginosar2019learning}, they used a combination of a L1 regression loss for the model training signal and an adversarial loss to ensure realistic motion. The learned body-motion-to-hands correlation was versatile enough for several use-cases, namely conversational hand gesture synthesis, single-view 3D hand-pose estimation and synthesizing missing hands in motion capture data and image-based pose estimation data.

\subsubsection{Control input} \label{subsubsec:control-input}
 Although control can take either linguistic or non-linguistic forms, it is distinct because it can convey the explicit design and execution intent of an animator. Multiple works in motion synthesis use control as an additional input either during the training phase or the inference phase of learning-based models (e.g. \cite{holden2017phase, ling2020character}). Typically, during training, the control signal is used to train the system to generate animations with certain biomechanical constraints such as posture, gait, etc. During inference, control may be introduced to impose style-related constraints \cite{smith2019efficient} or user input \cite{holden2017phase,henter2020moglow,ling2020character}. 
 
 In the context of conversational gesture, Alexanderson et al.~\cite{alexanderson2020style} trained a probabilistic model that generated spontaneous co-verbal gesture that was conditioned on control constraints such as wrist height, radial extent and handedness. However, the constraints were introduced at training time, meaning modeling a new constraint required re-training the entire model. Habibie et al.~\cite{habibie2022motion} provided a more flexible approach. They first learn a speech-to-gesture motion search through a kNN algorithm, and then refine the motion using conditional GAN. 
 Style control can be exerted at runtime by dynamically restricting the portion of the database that the kNN algorithm is run on, allowing style variation even within an extended utterance without the need to retrain.

 \new{Control can also be imposed by implicitly specifying the desired gestures by learning emergent prototypes of gesture shape or form.} Qian et al.\ \cite{qian2021speech} \new{explored this idea by learning} conditional vectors, so-called ``template vectors'', that could determine the general appearance and thus narrow the potential range of plausible gestures. Their framework took in audio and a zero initialized condition vector, through a 1D UNet-based autoencoder, in order to generate the corresponding gestures as 2D joint positions. During training, they periodically updated the condition vector, through back-propagation, using the gradients computed on the L1 regression loss between the generated and ground-truth gestures. They regularized the template vector space through the KL-divergence between the vectors and a normal distribution. \new{They also separately} pre-trained a VAE to reconstruct ground truth gestures and used the \new{resulting} latent space to encode \new{gestures} into template vectors. At test time, they sampled arbitrary template vectors, either learned through back-propagation or extracted by the pre-trained VAE, to generate diverse gestures. 
 
 Animators typically want to specify high-level style parameters to convey design intent e.g. energetic oratory gesticulations or subdued gestures to convey sadness. Additionally, it is desirable to specify the style once in the workflow and for the animation system to generate arbitrarily many motions for that specification. However, there is a gap between desired abstract design intent and existing deep-learning-based style control systems that tend to rely on biomechanical constraints such as wrist speed, radius or height \cite{alexanderson2020style,habibie2022motion}. Style specification is also not data efficient, requiring as many samples as the size of the training set for the model to learn a style \cite{alexanderson2020style,ahuja2020style}. \new{We conclude this section by discussing several works that} proposed approaches for data-efficient style specification \cite{ghorbani2022exemplar,ghorbani2022zeroeggs,fares2022zero,ahuja2022low}.
 
 Ghorbani et al.~\cite{ghorbani2022exemplar,ghorbani2022zeroeggs} proposed a framework that improves on high-level style portrayal by using exemplar motion sequences that demonstrate the intended stylistic expression of gesture motion.
 Their framework was able to efficiently extract style parameters in a zero-shot manner, only requiring  a single example motion and was able to generalize to example motions (and therefore styles) unseen during training. Fares et al.~\cite{fares2022zero} used an adversarial framework to learn a speaker-style encoder that could generate speaker-specific style embeddings from novel multimodal inputs -- audio, text and gesture pose -- not seen during the training phase. The framework generated co-speech gestures in a style that is either consistent with the original speaker in the audio or a different speaker, depending on the chosen style embedding. Ahuja et. al \cite{ahuja2022low} proposed an adversarial domain-adaptation approach for personalizing the gestures of a source speaker with plenty of data, with the style of a target speaker with limited data, using only 2 minutes of target training data. Given a model pretrained on a large co-speech gesture dataset, their framework could adapt the model's parameters using a smaller target dataset by modeling the cross-modal grounding shift, i.e., the change in distribution of speech-gesture associations, and the distribution shift in the target gesture space. The approach's ability to identify distributions shifts between the source and target domain for parameter updates, enabled the model to extrapolate to gestures in the target distribution without having seen them in the source distribution during pretraining.


\section{Key Challenges of Gesture Generation}

Animating co-verbal gestures is still a very challenging problem because gestures are spontaneous, highly idiosyncratic and non-periodic. Rule-based approaches generate well-formed gestures by leveraging recording motion, but are inflexible and lack gesture diversity. Additionally, the hand-designed rules are non-exhaustive and often prescriptive, and hence may not be reflective of gestures which occur naturally and spontaneously. Data-driven approaches improve on diversity and flexibility but tend to produce marginally natural gestures that appear more like well-timed hand waving, are not communicative and have little meaning. Although state-of-the-art systems employ speech and/or text information, they still do not handle semantic grounding of gestures properly, evidenced by gestures that seem to lack meaningful information when compared to the ground truth. Furthermore, due to the probabilistic nature of gestures, its idiosyncrasies, rich semantic content makes the evaluation process especially challenging and subjective. \new{In this section, we discuss the limitations of the current work and possible future directions in context of what we view as the key challenges of gesture generation, namely:
\begin{enumerate}
\item evaluation (in Section~\ref{subsec:challenge-evaluation}),
\item data (in Section~\ref{subsec:challenge-data}),
\item human-like gestures (in Section~\ref{subsec:challenge-humlike}),
\item multimodal grounding (in Section~\ref{subsec:challenge-grounding}), and
\item multimodal synthesis (in Section~\ref{subsec:challenge-multimodal}).
\end{enumerate}
}


\new{\subsection{Evaluation}
\label{subsec:challenge-evaluation}}
\new{Evaluation is of central importance to gesture generation,} both for developing co-speech gesture generation systems and for assessing their performance and capabilities in various aspects, as well as those of the field as a whole.
However, evaluating gestures is challenging due to the stochastic nature of gestures and the highly subjective nature of human gesture perception. A comprehensive review of evaluation practices in gesture generation can be found in \cite{wolfert2022review}.
\new{We recommend that readers consult that review regarding best practices, but also provide an overview of key open challenges in gesture evaluation here.}

\subsubsection{Subjective Evaluation}
\new{One important aspect to evaluate for gesture-generation systems is the human-likeness of the generated gestures}, which is measured and compared through human perceptual studies, often with comparable stimuli presented side by side as in e.g. \cite{jonell2021hemvip,kucherenko2021large,wolfert2021rate}. On the other hand, evaluating the \new{other aspects such as the} appropriateness and/or specificity of generated gestures in the context of speech and other multimodal grounding information (see Section~\ref{subsec:challenge-grounding}) is quite challenging, \new{especially since differences in the human-likeness of the motions being compared tends to interfere with perceived gesture appropriateness} (cf.\ the results in~\cite{kucherenko2021large}). To alleviate this challenge for appropriateness, a new evaluation paradigm of matched vs.\ mismatched gesture motion has recently been proposed \cite{jonell2020letsfaceit,rebol2021passing,yoon2022genea}.
In this setup, human participants are asked to choose between two motion clips that both were generated by the same system, and therefore have similar appearance and human-likeness, but where one clip is intended to be appropriate to the situation (e.g., the motion in it corresponds to the actual speech audio in the video) whereas the other is chosen at random (e.g., it was generated by feeding unrelated speech audio into the same system instead, and does not match the actual audio track).
The extent to which humans are able to identify the video that matches the situation can be used both to probe the strength of grounding in different modalities, and to assess gesture appropriateness for speech, rhythm, interlocutor behavior, etc., while controlling for human-likeness.
We expect this methodology to gain wider adoption and advance the state of the art in subjective assessment of different aspects of co-speech gestures.

\new{Another compelling area for future work is to evaluate gesture generation in actual interactions, since the ultimate goal of embodied conversational agents is to enhance human-computer communication and interaction. Initial studies \cite{nagy2021framework,he2022evaluating} have found that embodied agents that perform gestures generated by data driven models as opposed to performing no gestures, attract more attention from the audience. A larger attention span on a gesticulating agent is indicative of a more engaging communicative quality of gestures and opens doors to evaluating gesture generation in a more natural setting. Although the situated and time-demanding nature of such interactions, coupled with their reliance on many non-gesture components necessary to create interactivity (e.g.\ human wizards or automatic speech recognition, dialogue systems, and text-to-speech), make proper interactive evaluation challenging and seldom done, it is an important long-term goal for evaluations in the field.}
Given the difficulties in comparing different research papers in the field, we think that controlled, large-scale comparisons 
\cite{kucherenko2021large,yoon2022genea} with open data and materials are going to play an important role to develop the co-speech gesture field and its evaluation practices in the shorter term. This is similar to the role challenges have played in the development of text to speech \cite{king2014measuring} and the wide use of leaderboards and benchmarks across deep learning today.

\subsubsection{Objective Evaluation}
While subjective metrics from appropriately designed user-studies are the gold standard in co-speech gesture evaluation \cite{wolfert2022review}, they are expensive and time consuming, and thus lack scalability. There is therefore interest in objective metrics to automatically assess synthetic motion, for example its its human-likeness.
Objective metrics are useful to measure progress during model development in a heavy compute, data-driven learning setup. A natural metric is accuracy of prediction (i.e., how often the predicted position of a joint is within some tolerance of the joint position in a human motion capture clip), which is often called the Probability of Correct Keypoints (PCK). However, this quantity is often not indicative of performance due to the one-to-many nature of the gesture-generation problem. Two examples of human motion for the same speech might involve very different joint positions, and thus have low mutual agreement. \new{Measuring the mean squared error (MSE) between generated motion and human motion capture suffers from the same issue.}

Statistics of motion properties such as acceleration and jerk have been used as an alternative for quantifying and comparing generated gesture distributions \cite{kucherenko2021moving}, \new{but there is no compelling evidence that these metrics correlate with subjective assessments of motion human-likeness}.
To improve the measurement of distributional similarity of gestures, new objective quality metrics based on innovations from image processing, namely the Fr{\'e}chet Inception Distance (FID) \cite{heusel2017gans} and the Inception Score \cite{salimans2016improved}, were proposed in \cite{ahuja2020no,yoon2020speech} and \cite{ahuja2020style} respectively. Among these proposals, only \cite{yoon2020speech} computes the Fr{\'e}chet distance in a learned space. There has also been work in learning to estimate the \new{human-likeness} of gestures from databases of gesture motion and associated subjective ratings data \cite{he2022automatic}.
However, learning to predict human preference can be difficult even from relatively large training databases, as seen in similar research into predicting the subjective ratings of synthetic speech \cite{huang2022voicemos}.
A recent preprint \cite{kucherenko2023evaluating} used the data from the GENEA Challenge 2022 to compute correlations between subjective ratings and a number of objective metrics.
They found that almost none of the objective metrics displayed a statistically significant correlation with the human-likeness scores from the large user study, with the Fr{\'e}chet Gesture Distance being the one exception.

\new{Since the above approaches depend on motion data only, they can only give an indication of whether or not generated motion is statistically similar to the human motion capture in the database, but not how appropriate the motion is for the context in which it occurs (whether it is \emph{grounded} in that context). The methods can therefore not assess whether or not the motion is synchronized with the co-occurring speech, whether the motion is semantically relevant, etc. In general, unlike human-likeness, not many techniques have been proposed for objectively quantifying properties like gesture diversity or different kinds of motion appropriateness.}
One exception is the recent Semantic Relevance Gesture Recall (SRGR) metric from \cite{liu2022beat}, which proposes to quantify the semantic relevance of gesture by using semantic scores, annotated in the speech text data, to weight the probability of correct keypoints between the predicted and ground-truth gestures higher when the ground-truth gesture has a high semantic score.
\new{This is a step in the right direction for evaluating semantic appropriateness, but may suffer from the same issues as regular PCK due to the idiosyncratic, one-to-many nature of gesticulation.}
Given the impact that the Inception Score and the Fr{\'e}chet Inception Distance have had in driving progress in image generation, reliable metrics that estimate gesture human-likeness and especially appropriateness for e.g.\ the rhythm and semantics of co-occurring speech are an important continuing challenge, where recent and future innovations are likely to have significant impact on the field.

\subsection{Data}
\label{subsec:challenge-data}
Compared to machine-learning applications in text, speech, and images, gesture-generation is currently a data-limited field. 
A particular bottleneck is finger motion, which is difficult to capture accurately even through motion capture; cf.\ Table~\ref{table:datasets}. When finger motion is unreliable or unavailable, a possible mitigation might be to predict finger motion from other information, for example the rest of the body as in \cite{ng2021body2hands}.
In general, motion capture data is high quality, but laborious to capture, particularly when considering large scale data corpora.
Other issues arise due to the high variation in gesture behavior.  It can vary based on the individual, the environment, the number of people interacting, their emotional state and the topic of the conversation.
Some of this variation is grounded in information that cannot be effectively recorded because it, e.g., is internal to a speaker (such as their emotional state), or that is rarely captured, such as properties of the space in which an interaction is taking space.
But even if one were to capture or control for many of these these sources of variation, a great diversity in gesture behavior and realization would persist, which will be difficult to cover in any database we can record.


In the long term, if we can achieve sufficiently reliable 3D gesture extraction from monocular, in-the-wild online video, that will be a game-changer for the field of co-speech gesture generation.
It promises to have a transformative impact on both perceived authenticity and model capabilities, similar to how very large datasets for deep learning has powered recent advances in generative models for text and images, such as GPT-3~\cite{brown2020language}, DALL-E ~\cite{ramesh2021zero, ramesh2022hierarchical}, and Stable Diffusion ~\cite{rombach2022high}.
At present, works that study the use of in-the-wild data for gesture synthesis exist, for example \cite{yoon2019robots,ginosar2019learning,yoon2020speech,habibie2021learning,ahuja2020no}, but the quality of the data and the gestures do not yet amount to such a leap forward.

\subsection{Human-Like Gestures}
\label{subsec:challenge-humlike}
The most prominent research target in deep-learning-based co-speech gesture generation has long been perceptual quality.
This is similar to the focus on perceptual surface quality in other areas such as image generation ~\cite{karras2020analyzing, ramesh2021zero, ramesh2022hierarchical, rombach2022high} and speech synthesis ~\cite{wang2017tacotron, oord2016wavenet}.
One reason for this focus might be that perceptual surface quality is easier to estimate using standardized procedures, compared to quantities such as ``gesture appropriateness for speech''.
See, especially, the rapid quality improvements in the image-synthesis field, once reasonable objective metrics such as the Inception Score~\cite{salimans2016improved} and the Fr{\'e}chet Inception Distance~\cite{heusel2017gans} became available.

Just like deep generative methods in general have advanced greatly in recent years, there is strong evidence from large evaluations that the human-likeness of the best gesture-generation systems is improving as well \cite{kucherenko2021large,yoon2022genea}.
The better the visual quality of the avatar and greater range of expressive motion, the easier it should be to spot differences between natural and synthetic motion.
From this perspective, head motion (which only has three degrees of freedom) might for example be easier to make indistinguishable from human head motion, than it is to generate convincing arm and finger motion.
In this light, the achievement of GestureMaster \cite{zhou2022gesturemaster} in the GENEA Challenge
2022~\cite{yoon2022genea} is particularly noteworthy, since the synthesized upper- and full-body gestures produced by this model were rated higher than the original motion capture from the human speaker.
Although a very impressive result, this may partly be attributed to the
presence of some motion clips with motion-capture artifacts, especially for the fingers, that may reduce the perceived human-likeness of the notional human reference motion.

At the same time, even ``high quality'' gesture motion on a high-fidelity avatar is still judged as being far from human:
in the GENEA Challenge 2022~\cite{yoon2022genea}, neither the human motion capture nor the best performing system came near the rating of 100 that would correspond to being ``completely human-like''. More specifically, the median human-likeness of the best performing synthesis system were 69 for upper-body motion and 71 for full-body motion, with scores of 63 and 70 for human motion capture, respectively.
Our statement comes with several caveats.
Some of the gap up to a score of 100 might be attributable to shortcomings of motion capture when it comes to capturing the full range of human expression. For example, how an avatar moves and its lack of face, mouth, gaze and lip motion behavior can impact the visual qualities of the avatar.
Even in the case of speech synthesis, where recreating human behavior is as easy as playing back an audio recording, it is well known that humans tend to rate the human-likeness of high-quality recordings of human speech as around or below 4.5 on a 5 point scale; see for example the naturalness scores in the large and careful evaluation in \cite{ling2021blizzard}. Complete human-likeness may thus in practice be achieved at a score below the maximum on the any given ratings scale.
All that said, we believe that human-likeness can and will be improve further in the future, especially with more accurate motion capture and more lifelike avatars to display motion on.

As for the path that the gesture generation will take towards achieving new heights in human-likeness, we can look to history, and to other fields.
Data-driven generative modeling like~\cite{ginosar2019learning,ferstl2020adversarial, ahuja2020no,kucherenko2020gesticulator, yoon2019robots} took over as the state of the art in co-speech gesture generation with the advent of publicly available motion capture datasets suitable for training deep-learning architectures.
Since then, a variety of deep generative approaches have been applied (see Table~\ref{tab:dl-sota}), and human-likeness keeps improving \cite{kucherenko2021large,rebol2021passing,yoon2022genea}.
There is no doubt interesting work to come in applying recent diffusion models \cite{sohl2015deep,song2019generative,ho2020denoising}, already considered for general motion synthesis \cite{tevet2022human}, to gesture generation. While generated gestures from data-driven machine learning models are convincing, a lack of large scale gesture datasets currently limit the human-likeness of these approaches. Hence, in the short term, we may expect hybrid systems such as GestureMaster~\cite{zhou2022gesturemaster, ao2022rhythmic}
to be the leaders in human-like gesture generation.
Specifically, these are systems where machine-learning decides which general properties are needed of the gestures, but the actual gesture motion is primarily realized by assembling pre-recorded motion clips and frames, like in motion graphs \cite{lee2002interactive,kovar2002motion,arikan2002interactive,treuille2007near} and motion matching \cite{clavet2016motion}.
In the long-term, however, purely deep learning models are likely to take over. This would match the trajectory followed by text-to-speech synthesis, where hybrid systems once gave the best perceptual quality \cite{king2014measuring}, but pure deep-learning-based approaches trained on very large speech databases have recently taken the crown \cite{tan2021survey}.

\subsection{Multimodal Grounding}
\label{subsec:challenge-grounding}
Visually human-like gesticulation is not the only goal of gesture generation.
As discussed in the introduction to this article, a key goal with generating co-speech gestures is to facilitate communication, in much the same way as gestures enrich human communication.
This requires gestures that not only exhibit human-like movement on the surface but also are appropriately \emph{grounded} in the context of the interaction, so that they can contribute to it.
In more engineering-oriented terms, systems must take many relevant modalities as input, and make use of this information in an adequate way, to obtain synthetic gestures that can fulfill the same communicative roles as human gesticulation does.
It can be difficult to capture this information both in training data and at synthesis time, as well as to make meaningful use of it in the gesture generation.

\new{Grounding information can take many forms. Consequently, this section discusses challenges in grounding gesture-generation in a variety of relevant multimodal aspects (system inputs), beginning with aspects internal to the speaking agent, and then discussing grounding in other parties in the conversation as well as in the surrounding space.
More specifically, we cover grounding in
\begin{enumerate}
\item temporal information (Section~\ref{subsubsec:challenge-grounding-temporal});
\item semantic content (Section~\ref{subsubsec:challenge-grounding-semantic});
\item speaker identity, personality, emotion, and style (Section~\ref{subsubsec:challenge-grounding-style});
\item interlocutor behavior (Section~\ref{subsubsec:challenge-grounding-interlocutor}); and
\item spatial information (Section~\ref{subsubsec:challenge-grounding-spatial}).
\end{enumerate}
We also discuss some derived challenges posed by the often weak correlation between grounding information and the gesture motion (Section~\ref{subsubsec:challenge-grounding-correlation}), and how gestures may be grounded in the creative intent of a system designer (Section~\ref{subsubsec:challenge-grounding-intent}).}


\new{\subsubsection{Temporal Grounding}
\label{subsubsec:challenge-grounding-temporal}}
Gestures are temporal, which is a result of their correlation with a heavily temporal acoustic modality, along with the fact that they might depict occurrences or trace out paths or shapes over time.
The rhythmic nature of the gestures (i.e. beat gestures) in context of acoustic prosody has been studied heavily since the era of rule based gesture synthesis ~\cite{cassell1999speech, marsella2013virtual}. Fast forward to approaches with data-driven synthesis, some explicitly rely on extracted prosodic features \cite{ferstl2021expressgesture}, while others ~\cite{ginosar2019learning, ahuja2020style} learn implicit embeddings from acoustics which prosody is one of the key components. It seems clear that gesture production must be grounded in the rhythm of audio data, and appropriate beat gestures will be challenging to achieve from text transcriptions alone, without timing information \cite{kucherenko2022multimodal}.
Alternatively, both audio and gesture must be synthesized to have comparable rhythmic structure.


\new{\subsubsection{Semantic Grounding}
\label{subsubsec:challenge-grounding-semantic}}
Beyond the rhythmic nature of gestures, there is often a semantic meaning associated with the performed gesture.
\new{The small size of gesture-generation databases, and the complicated relationship and weak correlation between speech semantics and gesture form (see Section~\ref{subsubsec:challenge-grounding-correlation}), mean that it is unrealistic to expect systems to learn to generate semantically appropriate gestures driven by speech acoustics alone.}
Text, on the other hand, is a compact way to represent much of the semantic content behind co-speech gestures, and has been heavily studied since the era of rule-based gesture synthesis~\cite{cassell2001beat} as well as in data-driven synthesis~\cite{sadoughi2019speech, lee2021crossmodal, ahuja2020no,kucherenko2020gesticulator,yoon2020speech,zhou2022audio,liang2022seeg}. Current data-driven approaches typically attempt to gain semantic awareness by relying on deep-learning based language models trained on large amounts of text, such as~\cite{mikolov2013efficient, devlin2018bert}. Recent large language models based on large amounts of text \cite{brown2020language} have indeed been capable of generating text with surprisingly coherent semantics, suggesting that they can capture lexical meaning to a significant extent. While the inclusion of text has improved human perception of automatically generated gestures~\cite{kucherenko2020gesticulator, ahuja2020no, yoon2020speech, ao2022rhythmic}, it is still not trivial to measure the semantic content of gestures (see the discussion in Section~\ref{subsec:challenge-evaluation}). Hence, it is unclear how much (if any) of the improved human perception can be attributed to the semantic awareness created due to the use of language models, nor how much of the bottlenecks that exist may be removed with continuing progress in neural language models.
More broadly, there is a need for gesture synthesis models to perform better with regard to semantics.  Gesture is most powerful when it conveys information, and doing this effectively has been a challenge for most deep learning systems; cf.\ Figure \ref{fig:stefan_overview}.

\new{\subsubsection{Identity, Style, Emotion, and Personality}
\label{subsubsec:challenge-grounding-style}}
Co-speech gestures are idiosyncratic. The manifold of gestures performed by a speaker are not just a function of the content of the speech, but are also dependent on the identity, emotional state and the context of the speaker.
Generating personalized gestures based on speaker identity became possible with the influx of large scale multi-speaker datasets \cite{yoon2020speech, ahuja2020style}. Several GENEA Challenge 2022 \cite{yoon2022genea} systems also make use of speaker identity. A deeper analysis of the impact of speaker identity input \cite{kucherenko2022multimodal} shows that different speakers have different gesture-property prediction certainty, evoking even more interest in the idiosyncrasies of co-speech gestures. 
More recently, it was also shown that a short motion clip can be used for style control in ``zero-shot style adaptation''~\cite{fares2022zero,ghorbani2022exemplar,ghorbani2022zeroeggs}.  For many applications, it is desirable for the designer to be able to control the nature of the motion.  This goes beyond replicating idiosyncratic motion recorded of an individual to being able to specify novel characters.  We are far from having ways to author a character with an imagined personality for a particular application.

Apart from the speaker identity, the emotional or affective state of a speaker also impacts the gestures performed by them.
A striking example of this is the large range of expressive motion variation with the same lexical message explored in the Mimebot data \cite{alexanderson2017mimebot}.
Building emotionally aware embodied agents is a common research direction ~\cite{chowanda2016computational, sohn2018emotionally}. More recently, data-driven models have been explored where affective cues were learned using a dedicated encoder in an adversarial setup \cite{bhattacharya2021speech2affectivegestures} to imitate these patterns of affective behavior.
It is important to be able to drive these emotions in a way that is consistent with a character's personality and to be able to shift mood and emotion over time.
One way forward might be to leverage findings from the literature of gesture and motion perception, which has identified many useful properties of gesture motion that correlate with the perception of personality \cite{lippa1998nonverbal,koppensteiner2010motion,smith2017understanding} and emotion \cite{normoyle2013effect,castillo2019we}.
By changing these properties in synthesized gestures, we may exert some control over the perceived speaker personality and emotion ~\cite{alexanderson2020style, habibie2022motion}.
Again, speech synthesis provides an analogy, where it was recently shown that simple and easy additions of filler words and pausing can meaningfully and reliably be used to alter listeners' perception of speaker certainty ~\cite{kirkland2022s}.

\new{\subsubsection{Interlocutor-Aware Gestures}
\label{subsubsec:challenge-grounding-interlocutor}}
While non-verbal behavior is impacted by internal state of the speaker, the external context also guides the types of gestures a speaker might perform. In a dyadic conversation, the model must be aware of the behavior of the interlocutor while generating the relevant gestures~\cite{ahuja2019react,jonell2020letsfaceit,nguyen2022context, yang2020statistics}. This includes modeling appropriate listener behavior as well as speaker behavior.  Characters must modify their behavior to react to the content, mood and timing of interlocutors.  Characters must be able to be surprised, angered, pleased, etc. based on what their interlocutor may say.
Given the increasing availability of dyadic datasets with motion capture for both conversational parties, we expect to see more research in this direction in the next few years.

\new{\subsubsection{Spatially Aware Gestures}
\label{subsubsec:challenge-grounding-spatial}}
Even more generally, the context could also include spatial understanding of the environment. For example, the correctness of deictic gestures relies on the information about objects and directions in a scene. To carry communicative value, most of these gestures will therefore require access to visual and/or spatial information beyond what may be contained in the speech -- think about a phrase such as ``You need to go \emph{that} way'', which completely lacks information about which direction the system should point.  People also use spatial configurations in complex ways while gesturing, for example, placing ideas in a referential space in front of them and then referring to ideas by referring to the space they have been located in. While studies that involve external contexts are quite common for downstream tasks like navigation~\cite{savva2019habitat}, non-verbal behavior generation in multiple external contexts is up and coming~\cite{deichler2022towards, kucherenko2022multimodal} which makes it a promising research direction, if relevant data can be obtained.


\new{\subsubsection{Weak Correlations with Grounding Information}
\label{subsubsec:challenge-grounding-correlation}}
Let's imagine that we have access to all the variables discussed thus far that impact the dynamics of co-speech gestures, such as acoustics, text, speaker identity, emotional state, and external contexts. Further imagine that we are able to gather large-scale datasets with all these variables, which
is unlikely to ever happen due to the combinatorical explosion of possible combinations of different factors.
Would having this rich input information and broad data coverage be sufficient to confidently predict the specific co-speech gestures that a given speaker will perform? The best we can likely say is ``Maybe!'' While large scale datasets may enable us to minimize our epistemic uncertainty about gesticulation, it is unclear how significant the stochasticity is, i.e.\ aleatoric uncertainty, of these gestures will be. The situation is analogous to the problem of prosody in text-to-speech, where there can be many possible acoustic realizations and intonation contours for the same lexical input~\cite{watts2015sentence, luong2017adapting, wang2018style}. Significant variation persists even when a speaker is asked to read the same text several times under exactly the same circumstances~\cite{henter2014measuring}.
To handle ambiguity in gesture realization, it is compelling to consider probabilistic models, since they can ``hallucinate'' the missing information and stochastic components of non-verbal behavior, as a way to resolve the one-to-many problem for motion and gesture generation \cite{henter2020moglow,alexanderson2020style}.

\new{\subsubsection{Grounding Gestures in Creative Intent}
\label{subsubsec:challenge-grounding-intent}}
Gesture authoring enables an animator or system creator to design and edit motion, e.g.\ making a character appear less nervous or stressed, thus grounding the animation within the designer's creative intent. Typically, animation design intent is captured through key-framing or motion capture. However, these approaches are difficult to scale for nonverbal behavior because the former requires specialized animation skills, while the latter requires expensive camera setups and laborious post-processing. Automatic gesture generation approaches in part solve the scalability issue by the ability to generate abundant motion data, but they struggle with high-level control. For instance, attempts at handling control either bake in mechanistic, low-level control signals like wrist height, wrist velocity, and radial extent \cite{alexanderson2020style}, or they generate gestures that deviate from the intended control specifications \cite{habibie2022motion}. Moreover, in multi-speaker scenarios, they are unable to capture the variability of different speakers' gesticulation, and cannot distinguish between gesture types used in a certain scenario (e.g.\ deictic gestures for a lecturer in front of display) from gesture style differences between speakers \cite{ahuja2020style}. Yoon et al.~\cite{yoon2021sgtoolkit} recently proposed an innovative approach to this challenge: an authoring toolkit that balances gesture quality and authoring effort. The toolkit combines automatic gesture generation using a GAN-based generative model \cite{yoon2020speech} and manual controls. The generative model first produces a gesture sequence from speech input, and animator can interactively edit the motion through low-level pose control and coarse-level style parameters. We think similar gesture authoring approaches that maximize design intent and gesture quality, while minimizing authoring effort will be important for grounding nonverbal behavior within the animator's creative intent.

\subsection{Multimodal Synthesis}
\label{subsec:challenge-multimodal}
Human communicative behavior is not only grounded in multiple modalities and information streams, but is also expressed through multiple modalities.
A complete virtual agent agent will need to listen, observe, decide, speak, and move.
On the generation side, verbal behavior generation is considered separate from non-verbal behavior, and the generation of non-verbal behavior is in turn typically broken into several smaller sub-problems treated in isolation.
Head motion might be treated separately from lip motion, facial expression, and gaze; finger motion might be treated separately from arm motion; and lower-body motion might be separated from the motion of the upper body.
A long-term goal would be to bring these sub-problems together, to create more coherent synthetic behavior with a wider range of possible expressions, and eventually unify the synthesis of these expressions with verbal behavior generation. Recent work has explored learning full-body gesture motion (including the head and the lower body), e.g.\ \cite{alexanderson2020style} and the submissions to the full-body tier of the GENEA Challenge 2022 \cite{yoon2022genea}.

Another line of work has considered training verbal (text-to-speech) and non-verbal (speech-to-gesture) synthesis systems on the same data \cite{alexanderson2020generating} and, subsequently, merging them into one single network that generates both speech audio and gesture motion \cite{wang2021integrated}.
Given the strides that have been made in generating convincing speech audio from text \cite{tan2021survey}, adapting successful text-to-speech methods to simultaneously generate both acoustics and joint rotations, as was done in \cite{wang2021integrated}, seems like a compelling direction for future work.
This not only brings advantages in terms of modeling efficiency (the gesture-generation systems will possess information about, e.g.\ prosodic prominence without having to learn to extract that information from speech audio), but also more closely resembles models of human communication such as the growth-point hypothesis ~\cite{mcneil1992handmind}, and could enable gestures that not only complement but, as in Kendon's continuum (see Figure \ref{fig:kendonscont}), replace or augment speech with novel information. This may require even deeper representations of communicative intent, as approaches that generate gesture based on text and/or audio are restricted to redundant gestures, but gesture that is non-redundant with the spoken audio is a key part of human behavior.

\section{Broader Impact}
High quality gesture synthesis can advance a range of applications by allowing computational systems to leverage nonverbal communication.  This can allow more natural and fluid communication of both functional and affective information, which will prove useful in a range of assistive applications, employing both agents and robots. These include tutors, rehabilitation trainers, relational agents for health and eldercare, and personal assistants. They can also support richly interactive entertainment experiences in which you can have meaningful interactions with virtual characters.

The development of the technology also raises potential ethical issues which must be given careful consideration.  Some of the issues are common to many deep learning approaches that involve human data. For instance, what kind of bias is in the data that is used?  Does it represent the full range of human nonverbal behavior, or only specific language groups, ethnicities and social strata?  Will people using these models take care to match the input data with the desired output representation or will the data be mismatched, using the wrong gender, ethnicity, age, etc. on synthesized characters?  What are the ownership rights associated with data that may be scraped from a web source?  Do you own your gesture style?  How can consent be obtained for online data?

The technology could also make it easier to generate deepfakes, i.e., synthetic media that mimics the likeness of real people, especially of politicians and other public figures that have a lot of video data online. Prominent examples include photorealistic lip motion from audio \cite{suwajanakorn2017synthesizing}, real time facial expression re-enactment \cite{thies2016face2face} and talking-head video synthesis \cite{wang2021facevid2vid}. The technology can be adapted to create synthetic nonverbal motion for nefarious purposes such as political propaganda, financial fraud and fake news. Moreover, a more unique consideration for nonverbal behavior results from people's tendency to entrain to their interlocutors.  If they entrain to synthetic models they may interact with, does this have any impact on their own behavior?  It is important for both researchers and developers of this technology to devise ways to mitigate these risks, such as those suggested in \cite{horvitz2022horizon}.


\section{Conclusion}
This paper summarizes the history of gesture generation, from early work on rule-based systems to the explosion of recent work using deep learning approaches. Deep learning approaches have employed a range of input, including text, audio and various control signals, and used a wide set of architectures.  Most systems have focused on monologue generation, but work is beginning to explore dialog and richer notions of context.  Despite substantial progress, the field is still young and there are very significant challenges to solve. These include better datasets, improved subjective and objective evaluation practices, higher quality motion, producing more meaningful gestures, adequately addressing the stochasticity of gesture, providing adequate control over the output and matching the rich set of grounding that supports human gesture, from multi-person interaction to adequately representing the spatial context of the conversation. There is much exciting work to come. 

\section*{Acknowledgments}
S.\ N.\ was partially supported by an IBM PhD fellowship award. G.\ E.\ H.\ and T.\ K.\ were partially supported by the Knut and Alice Wallenberg Foundation, both through Wallenberg Research Arena (WARA) Media and Language -- with in-kind contribution from the Electronic Arts (EA) R\&D department, SEED -- and through the Wallenberg AI, Autonomous Systems and Software Program (WASP). S.\ N.\ and M.\ N.\ were partially supported by the National Science Foundation on grant IIS 2232066.

The authors are grateful to Michael Kipp for Figure \ref{fig:gesturetree}, Stefan Kopp for Figure \ref{fig:stefan_overview}, and to Konrad Tollmar and anonymous reviewers for reviewing the manuscript.

\bibliographystyle{eg-alpha-doi} 
\bibliography{main}    
\end{document}